\documentclass[11pt]{amsart}
\usepackage{a4}
\usepackage{amsthm,amssymb}
\usepackage[english]{babel}
\newcommand{\RR}{\mathbb{R}}

\newcommand{\NN}{\mathbb{N}}

\renewcommand{\SS}{\mathbb{S}}
\newcommand{\cA}{\mathcal{A}}

\newcommand{\cC}{\mathcal{C}}
\newcommand{\cE}{\mathcal{E}}

\newcommand{\cL}{\mathcal{L}}

\newcommand{\cO}{\mathcal{O}}
\newcommand{\cS}{\mathcal{S}}
\newcommand{\cT}{\mathcal{T}}

\renewcommand{\r}{\right}
\renewcommand{\l}{\left}
\newcommand{\la}{\langle}
\newcommand{\ra}{\rangle}
\newcommand{\fs}{\mathcal{G}}
\newcommand{\dd}{\mathrm{d}}
\newcommand{\cG}{c^\Gamma}
\newcommand{\mG}{m_\Gamma}
\newcommand{\Mg}{M_\gamma }
\newcommand{\tg}{\tilde \gamma}
\newcommand{\alp}{{\theta_\cL}}

\newcommand{\dist}{{\ensuremath{\mathrm{dist}}}}
\newcommand{\Hag}{H_{\alpha\sigma_\Gamma}}

\newcommand{\be}{\begin{equation}}
\newcommand{\ee}{\end{equation}}
\newcommand{\bea}{\begin{eqnarray*}}
\newcommand{\eea}{\end{eqnarray*}}
\newtheorem{thm}{Theorem}[section]
\newtheorem{lem}[thm]{Lemma}
\newtheorem{prp}[thm]{Proposition}
\newtheorem{cor}[thm]{Corollary}
\theoremstyle{definition}
\newtheorem{dfn}[thm]{Definition}
\newtheorem{rem}[thm]{Remark}

\newcommand{\Hm}[1]{\leavevmode{\marginpar{\tiny%
$\hbox to 0mm{\hspace*{-0.5mm}$\leftarrow$\hss}%
\vcenter{\vrule depth 0.1mm height 0.1mm width \the\marginparwidth}%
\hbox to
0mm{\hss$\rightarrow$\hspace*{-0.5mm}}$\\\relax\raggedright #1}}}

\begin{document}
\title[Lower bounds on the spectral gap]{Lower bounds on the lowest spectral gap of singular potential
Hamiltonians}
\author[S.~Kondej]{Sylwia Kondej}

\address[S.~Kondej]{Institute of Physics,\, University of Zielona Gora,\, ul.~Prof.~Z.~Szafrana 4a,\, Zielona Gora,\, Poland}
%\curraddr{Fakult\"at f\"ur Mathematik, \, 09107 \ TU\, Chemnitz,\, Germany}
\email{skondej@proton.if.uz.zgora.pl}
\author[I.~Veseli\'c]{Ivan Veseli\'c}
\address[I.~Veseli\'c]
{Fakult\"at f\"ur Mathematik, 09107\, TU\, Chemnitz \& Emmy-Noether-Programme of the DFG,  Germany}

\urladdr{www.tu-chemnitz.de/mathematik/schroedinger/members.php}
\thanks{To appear in slightly different form in \emph{Ann. Henri Poincar\'e.}}

%\date{\today, \currenttime h, \jobname.tex}

 \subjclass[2000]{}

\begin{abstract}
We analyze Schr\"odinger operators whose potential is given by a
singular interaction supported on a sub-manifold of the ambient
space. Under the assumption that the operator has at least two
eigenvalues below its essential spectrum we derive estimates on
the lowest spectral gap. In the case where the sub-manifold is a
finite curve in two dimensional Euclidean space the size of the
gap depends only on the following parameters: the length, diameter
and maximal curvature of the curve, a certain parameter measuring
the injectivity of the curve embedding, and a compact sub-interval
of the open, negative energy half-axis which contains the two
lowest eigenvalues.
\end{abstract}

\maketitle
\let\languagename\relax

\begin{center}
\emph{Dedicated to Kre\v simir Veseli\'c on the occasion of his 65$^{\text th}$ birthday.}
\end{center}

\bigskip

%\thispagestyle{myheadings}
%\markright{To appear in \emph{Ann. Henri Poincar\'e}. \hfill}

\section{Model and results}

This paper studies quantum Hamiltonians with singular potentials,
also called singular interactions. This kind of perturbations are
particularly important for nanophysics because they model
``leaky'' nanostructures. More precisely, the Hamiltonians
describe nonrelativistic quantum particles which are confined to a
nanostructure, e.g.~a thin semiconductor, with high probability,
but still allowed to tunnel in and through the classically
forbidden region.

An idealization of this situation is a $d$-dimensional quantum
system with the potential supported by a finite collection of
sub-manifolds $\Gamma \in \RR^d$ whose geometry is determined by
the semiconductor structure. In general the different manifolds in
the collection $\Gamma $ may have different dimensions. The
corresponding Hamiltonian may be formally written as
\begin{equation}
-\Delta - \alpha \delta(x-\Gamma)\,,
\end{equation}
where $\alpha >0$ denotes the coupling constant.

Various results concerning the spectrum of Hamiltonians with
singular pertubations were already obtained, for instance, in
\cite{BrascheNEKS-94}, and more recently in \cite{ExnerI-01},
\cite{ExnerY-02b}, \cite{ExnerY-03}, \cite{ExnerK-02},
\cite{ExnerK-03}. However almost nothing is known about gaps
between successive eigenvalues. Some estimates for spectral gaps
can be recovered from the results given in
\cite{ExnerY-02b},\cite{ExnerK-03} but only for the strong
coupling constant case, i.e. $\alpha \to \infty$. The aim of this
paper is to make progress in this field and obtain lower bounds
for the first spectral gap $E_1 -E_0$, where $E_1 \,, E_0$ are the
two lowest eigenvalues.

Since singular potentials are a generalization of regular ones let
us give a brief review of some facts known for the latter before
formulating our results. It is well known that the double well
potential with widely separated minima gives rise to eigenvalues
which tend to be grouped in pairs. The classical result by Harrell
(see \cite{Harrell-80}) shows that the magnitude of splitting is
exponentially small with respect to the separation parameter,
i.e.~the distance between the wells.
This leads naturally to the question whether for more general %class of
potentials $ V$ one can obtain bounds for eigenvalue splittings in
terms of the geometry of $V$ and a spectral parameter at or near
the eigenvalues in question. The problem was studied by Kirsch and
Simon in \cite{KirschS-85, KirschS-87}. It was shown that for one
dimensional Schr\"odinger operators $-\Delta +V$, where $V$ is a
smooth function supported on a set $[a,b]$, the eigenvalue gaps
can be bounded in the following way
\begin{equation}
\label{e-KirschS} E_n -E_{n-1} \geq \pi \lambda^2 e^{-\lambda
(b-a)}\,, \quad n \in \NN
\end{equation}
where
\[
\lambda=\max_{E\in \,]E_n\,, E_{n-1}[\, ; x\in \,]a,b[\,
}|E-V(x)|^{1/2},
\]
cf.~\cite{KirschS-85}. For the multi-dimensional case an
exponential lower bound for the spectral gap $E_1 -E_{0} $ was
found in \cite{KirschS-87}.

Our main result can be considered as the analog for singular
potentials of the results in \cite{KirschS-87}. Thus we return to
the main topic of this paper and ask the question: can one find
exponential lower bounds for the eigenvalues splittings for
Schr\"odinger operators with singular interactions? We address
this question for a two dimensional system with a potential
supported by a finite curve (or more generally finitely many
disconnected curves). The desired lower bound is expressed in
terms of geometric properties of $\Gamma $. A crucial role is
played by the diameter $2R$ of $\Gamma$.
\begin{itemize}
\item
The main aim of this paper is to show the following lower bound
(see Theorem \ref{t-spgaps})
\[
E_1 -E_0 \geq \kappa_1^2 \mu_{\Gamma ,\alpha }(\rho ,
\kappa_0)\mathrm{e}^{-C_0\rho}\,, \text{ with } \rho:= \kappa_0R
\]
where $\kappa_i =\sqrt{-E_i}$ and $C_0$ is a constant. The
dependence of the function $\mu_{\Gamma ,\alpha }$ on geometric
features of $\Gamma$ is given explicitely in equation
\eqref{defmu}.
\end{itemize}
To prove the above estimate we establish some auxiliary results
which, in our opinion, are interesting in their own right. They,
for example, concern the
\begin{itemize}
\item
generalization to singular potentials of techniques developed in
\cite{DaviesS-84}, \cite{KirschS-87} to estimate the first
spectral gap,
\item
analysis of the behaviour of eigenfunctions: exponential decay,
localization of maxima and nodal points,
\item
estimates for gradients of eigenfunction, in particular near the
support of the singular potential.
\end{itemize}
The last mentioned point concerns a step in our strategy which is
very different from the route taken in \cite{KirschS-87}. There,
in fact, a gradient estimate of eigenfunctions is derived relying
on the assumption that the potential is bounded --- a situation
quite opposite to ours.
\bigskip

The paper is organized as follows. In Section \ref{s-general} we
present some general facts about Hamiltonians with singular
potentials. In Section \ref{s-general-gap} we adapt to the
singular potential case an abstract formula for the first spectral
gap which was derived in \cite{DaviesS-84}, \cite{KirschS-87} for
regular potentials.

In Section \ref{s-curve} we specialize to the case where the
support of the potential is a finite curve $\Gamma$ in a
two-dimensional Euclidean ambient space. In this situation we
derive our most explicite lower bound on the first spectral gap in
terms of geometric parameters of $\Gamma$. The proof of this
result is contained in the three last sections.

Section \ref{s-pointwise} contains several estimates on the
pointwise behaviour of eigenfunctions. In Section \ref{s-gradient}
we establish upper and lower bounds on gradients of
eigenfunctions. Special attention and care are given to the
behaviour near the support of the singular interaction. A
technical estimate is deferred to Appendix \ref{s-appendix}.
Section \ref{s-Outlook} is devoted to the discussion of our results
and of some open questions.

{\small\textbf{Acknowledgment} It is a pleasure to thank David
Krej\v{c}i\v{r}\'{i}k for comments on an earlier version of this
paper. S.K. is grateful for the hospitality extended to her at the
Technische Universit\"at Chemnitz, where the most of this work was
done. The research was partially supported by the DFG under grant
Ve 253/2-1 within the Emmy-Noether-Programme.}

\section{Generalized Schr\"odinger operators}
\label{s-general}

We are interested in Hamiltonians with so called singular
perturbations. In general, this kind of perturbation is localized
on a set of Lebesgue measure zero. In this paper we consider more
specifically operators with an interaction supported on an
orientable, compact sub-manifold $\Gamma \subset \RR^d$ of class
$C^2$ and codimension one. The manifold $\Gamma $ may, but need
not, have a boundary.
\medskip

The Hamiltonian with a potential perturbation supported on $\Gamma $ can be
formally written as
\begin{equation} \label{e-delta}
-\Delta - \alpha \delta(x-\Gamma)\,,
\end{equation}
where $\alpha>0 $ is a coupling constant.

To give (\ref{e-delta}) a mathematical meaning we have to construct
the corresponding selfadjoint operator on $L^2:=L^2(\RR^d)$. The
scalar product and norm in $L^2 $ will be denoted by $(\cdot ,
\cdot)$ and $\|\cdot \|$, respectively. Let us consider the Dirac
measure $\sigma_\Gamma $ in $\RR^d$ with support on $\Gamma$, i.e. for
any Borel set $G\subset \RR^d$ we have
\[
\sigma_\Gamma (G):= s_{d-1}(G\cap \Gamma )\,,
\]
where $s_{d-1}$ is the $d-1$ dimensional surface measure on
$\Gamma $. It follows from the theory of Sobolev spaces that the
trace map
\[
I_{\sigma_\Gamma} \colon W^{1,2}\to L^2(\sigma_\Gamma )\,, \quad \text{ where }\quad
W^{1,2}:= W^{1,2}(\RR^d )\,,\,\, L^2(\sigma_\Gamma ):= L^2(\RR^d ,\sigma_\Gamma
)
\]
is a bounded operator. Using the trace map we construct the
following sesquilinear form
\begin{equation} \label{form}
\cE_{\alpha\sigma_\Gamma}(\psi,\phi )=\int_{\RR^d} \nabla \psi (x) \nabla
\overline \phi (x)\mathrm{d}x-\alpha \int_{\RR^d} (I_{\sigma_\Gamma}
\psi)(x) (I_{\sigma_\Gamma} \overline \phi )(x) \mathrm{d}\sigma_\Gamma (x)\,,
\end{equation}
for $\psi\,,\phi\in W^{1,2}$. From Theorem 4.1 in
\cite{BrascheNEKS-94} we infer that the measure $\sigma_\Gamma $
belongs to the generalized Kato class, which is a natural
generalization of the notion of Kato class potentials. In
particular, for such a measure and an arbitrary $a>0$ there exists
 $b_a<\infty$ such that
\[
\int_{\RR^d} |(I_{\sigma_\Gamma} \psi)(x) |^2 \mathrm{d}\sigma_\Gamma (x)
\leq
a\|\nabla \psi \|^2 +b_a\|\psi\|^2\,.
\]
This, in turn, implies that the form $\cE_{\alpha\sigma_\Gamma}$ is
closed. Consequently there exists a unique selfadjoint operator
$H_{\alpha\sigma_\Gamma}$ acting in $L^2$ associated to
$\cE_{\alpha\sigma_\Gamma}$. This operator $\Hag $ gives a precise
meaning to the formal expression (\ref{e-delta}).

\begin{rem}\label{r-bc}
Using an argument from \cite{BrascheNEKS-94} we can define the
operator $\Hag $ by appropriate selfadjoint boundary conditions on
$\Gamma$. Denote by $\mathrm{n}\colon \Gamma \to \SS^d$ a global
unit normal vectorfield on $\Gamma$. Let $D(\tilde{H}_{\alpha
\sigma_\Gamma})$ denote the set of functions
\[
\psi\in C(\RR^d)\cap W^{1,2}(\RR^d) \cap
C^\infty(\RR^d\setminus\Gamma) \cap W^{2,2}(\RR^d\setminus\Gamma)
\]
which satisfy \footnote{In \cite{BrascheNEKS-94} the notation $-\frac{\partial }{\partial n_{-}}$
is used for what we denote by $\partial_\mathrm{n}^{-}$.}
\begin{align}
\label{e-boundcond}
\partial_\mathrm{n}^{+} \psi (x) + \partial_\mathrm{n}^{-} \psi (x)
= - \alpha \psi(x) \quad \quad \text{ for } x \in \Gamma \,,
\intertext{ where } \nonumber
\partial_\mathrm{n}^{+} \psi (x)
:= \lim_{\epsilon \searrow 0} \frac{\psi(x+\epsilon
\mathrm{n}(x))-\psi(x)}{\epsilon}
\\ \nonumber
\partial_\mathrm{n}^{-} \psi (x)
:= \lim_{\epsilon \searrow 0} \frac{\psi(x-\epsilon
\mathrm{n}(x))-\psi(x)}{\epsilon}\,.
\end{align}
By Green's formula we have for $\psi,\phi \in D(\tilde{H}_{\alpha
\sigma_\Gamma })$
\[
-\int_{\RR^2} (\Delta\psi(x)) \, \overline{\phi(x)} \, \dd x =
\cE_{\alpha\sigma_\Gamma} (\psi,\phi)\,.
\]
Using this equation we can conclude exactly as in Remark 4.1 of
\cite{BrascheNEKS-94} that the closure of $-\Delta$ with domain
$D(\tilde{H}_{\alpha \sigma_\Gamma})$ is the selfadjoint operator
$H_{\alpha \sigma_\Gamma }$.

It can be immediately seen from formula \eqref{e-boundcond} that
the opposite choice of the orientation of the manifold $\Gamma$
does not change the boundary condition. It is useful to note that
the eigenfunctions of $\Hag $ belong to $C(\RR^n)\cap
C^\infty(\RR^n\setminus\Gamma)$, cf.\cite{ExnerY-03}.
\end{rem}

% \begin{modtiny}
% \begin{rem}\label{tmp-r-bc}
% From the discussion in Section 4 of \cite{BrascheNEKS-94} we
% conclude that functions $\psi\in C(\RR^n)\cap
% C^\infty(\RR^n\setminus\Gamma)$ which are in the operator domain
% of $\Hag$ satisfy the following boundary conditions at the curve
% $\Gamma$. (In fact, in \cite{BrascheNEKS-94} the authors consider
% sub-manifolds without boundaries, however the argument, relaying
% on the Green formula can be also repeated for the sub-manifolds
% with boundaries). Denote by $\mathrm{n}\colon \Gamma \to \SS^d$
% the global unit normal vectorfield on $\Gamma$. Then $\psi$
% satisfies for $x\in \Gamma$
% \begin{align}
% \label{tmp-e-boundcond}
% \partial_\mathrm{n}^{+} \psi (x) + \partial_\mathrm{n}^{-} \psi (x) = - \alpha \psi(x) \quad \quad
% \text{ where }
% \\ \nonumber
% \partial_\mathrm{n}^{+} \psi (x):= \lim_{\epsilon \searrow 0} \frac{\psi(x+\epsilon \mathrm{n}(x))-\psi(x)}{\epsilon}
% \\ \nonumber
% \partial_\mathrm{n}^{-} \psi (x):= \lim_{\epsilon \searrow 0} \frac{\psi(x-\epsilon \mathrm{n}(x))-\psi(x)}{\epsilon}\,.
% \end{align}
% It can be immediately seen from the formula that the opposite
% choice of the orientation of the manifold $\Gamma$ does not change
% the boundary condition.
%
% \Hm{They assume that sub-manifold 'closed';
% please check if you can whether the argument works for
% sub-manifold with boundaries. They quote Agmon's lectures, which I
% do not have.}
% It is useful to note that the eigenfunctions of $\Hag $ belong to
% $C(\RR^n)\cap C^\infty(\RR^n\setminus\Gamma)$,
% cf.\cite{ExnerY-03}.
% \end{rem}
% \end{modtiny}

\begin{rem}
The manifold $\Gamma$ may have several components. We will provide
proofs for our results only in the case that $\Gamma$ is
connected. The modification for the general case consists
basically in the introduction of a new index numbering the
components. See also the Remark \ref{r-RandL}.
\end{rem}

\begin{dfn}[Resolvent of $\Hag$]
Since we are interested in the discrete spectrum of $\Hag$ we
restrict ourselves to values in the resolvent set with negative
real part.

For $\kappa >0$ denote by $R^\kappa :=(-\Delta +\kappa^2 )^{-1}$
the resolvent of the ``free'' Laplacian. It is an integral
operator for whose kernel we write $G^k (x-x')$. Furthermore
define $R^\kappa_{\sigma_\Gamma ,\mathrm{d}x}$ as the integral operator
with the same kernel but acting from $L^2$ to $L^2 (\sigma_\Gamma )$. Let
$R^\kappa_{\mathrm{d}x, \sigma_\Gamma }$ stand for its adjoint, i.e.
$R^\kappa_{\mathrm{d}x, \sigma_\Gamma }f=G^\kappa\ast f\sigma_\Gamma $ and
finally we introduce $R^\kappa_{\sigma_\Gamma, \sigma_\Gamma}$ defined by
$G^\kappa$ as an operator acting from $L^2 (\sigma_\Gamma)$ to itself. In
the following theorem we combine several results borrowed from
\cite{BrascheNEKS-94} and \cite{Posilicano-04, Posilicano-01}.
\end{dfn}

\begin{thm}\label{resolvent}
\noindent (i) There is a $\kappa_0 >0$ such that operator
$I-\alpha R^{\kappa }_{\sigma_\Gamma, \sigma_\Gamma } $ in $L^2 (\sigma_\Gamma)$ has
a bounded inverse for any $\kappa \geq \kappa_0$.

\noindent (ii) Assume that $I-\alpha R^{\kappa}_{\sigma_\Gamma, \sigma_\Gamma}$
is boundedly invertible. Then the operator
\[
R_{\alpha\sigma_\Gamma }^\kappa =R^\kappa + \alpha R^\kappa_{\mathrm{d}x,
\sigma_\Gamma } (I-\alpha R^{\kappa }_{\sigma_\Gamma , \sigma_\Gamma }
)^{-1}R^\kappa_{\sigma_\Gamma ,\mathrm{d}x}
\]
maps $L^2 $ to $L^2 $, $-\kappa^2 \in \rho (\Hag)$ and $R_{\alpha\sigma_\Gamma }^\kappa
=(\Hag +\kappa^2 )^{-1}$.

\noindent (iii) Suppose $\kappa >0$. The number $-\kappa^2$ is an
eigenvalue of $\Hag$ iff $\mathrm{ker}( I-\alpha R^{\kappa }
_{\sigma_\Gamma , \sigma_\Gamma })\neq \{0\}$. Moreover,
\[
\dim \ker (\Hag +\kappa^2 )= \dim \ker ( I-\alpha R^{\kappa }
_{\sigma_\Gamma , \sigma_\Gamma }).
\]
\noindent (iv) Assume $-\kappa^2$ is an eigenvalue of $\Hag$. Then
for every $w_\kappa \in \ker ( I-\alpha R^{\kappa }_{\sigma_\Gamma ,
\sigma_\Gamma })$ the function defined by
\begin{equation} \label{e-ef}
\psi_\kappa := R^{\kappa }_{\mathrm{d}x ,\sigma_\Gamma }w_\kappa \,.
\end{equation}
is in $D(\Hag)$ and satisfies $\Hag \psi_\kappa =
-\kappa^2\psi_\kappa $.
\end{thm}

Combining the statements (iii) and (iv) of the above theorem we
get the equality
\begin{equation}\label{e-relationpsiw}
\alpha I_{\sigma_\Gamma} \psi_\kappa =w_\kappa \,,
\end{equation}
which will be useful in the sequel, more precisely in equation
\eqref{e-norm3}.

\begin{rem}[Some facts about the spectrum of $\Hag$]
\label{r-spectrum}

Since the perturbation is supported on a compact set the essential
spectrum of $\Hag$ is the same as for free Laplacian, i.e.
\[
\sigma_{\mathrm{ess}} (\Hag)=[0,\infty[\,,
\]
cf.~\cite{BrascheNEKS-94}. From \cite{BrascheNEKS-94,ExnerY-03}  
we infer that $\Hag$ has nonempty discrete spectrum if $d=2$ and $\alpha $ is
positive. For $d\geq 3$ there is a critical value $\alpha_c > 0$
for the coupling constant such that the discrete spectrum
of $\Hag$ is empty if and only if $\alpha \le  \alpha_c $. 
The discrete spectrum has been analyzed in various papers (see
\cite{ExnerY-02b,ExnerY-03,ExnerY-04} and \cite{Exner-03}). It
was shown, for example, that for a sub-manifold $\Gamma $ without
boundary we have the following asymptotics of the $j$-th
eigenvalue of $\Hag$ in the strong coupling constant limit
\begin{equation} \label{e-asymev}
E_j (\alpha ) =-\frac{\alpha^2}{4} +\mu_j + \cO\left(
\frac{\log \alpha }{\alpha } \right) \quad \mathrm{as}\quad \alpha
\rightarrow \infty\,,
\end{equation}
where $\mu_j$ is the eigenvalue of an appropriate comparison
operator. This operator is determined by geometric properties of
$\Gamma $, i.e.~its metric tensor. In the simplest case, when
$\Gamma \subset \RR^2$ is a closed curve of length $L$ determined
as the range of the arc
length parameterization $[0,L] \ni s\mapsto \gamma (s)\in \RR^2$,
the comparison operator takes the form
\[
-\frac{d^2}{ds^2}-\frac{\varkappa (s)^2}{4}\,:\, D \left(
\frac{d^2}{ds^2} \right)\rightarrow L^2 (0,L)\,,
\]
where $\frac{d^2}{ds^2}$ is the Laplace operator with periodic
boundary conditions and
$\varkappa :=|\gamma''|\colon[0,L]\to \RR$
is the curvature of $\Gamma$ in the parameterization
$\gamma \colon [0,L]\to \RR^2$.
\medskip

If the curve $\Gamma $ is not closed analogous
asymptotics as (\ref{e-asymev}) can be proven. However now only
upper and lower bounds on $\mu_j$ can be established, namely
\[
\mu_j^ N \leq \mu_j \leq \mu_j^D\,,
\]
where $\mu_j^ N , \mu_j^ D$ are eigenvalues corresponding to
Neumann, respectively Dirichlet boundary conditions of a
comparison operator.
\end{rem}

\section{The lowest spectral gap for singular perturbations}
\label{s-general-gap}

The aim of this section is to give general formulae for the first
spectral gap of $\Hag $. Following the idea used for regular
potentials (see for example (\cite{DaviesS-84}, \cite{KirschS-87})
we will introduce a unitary transformation defined by means of the
ground state of $\Hag$.

Assume that there exists an eigenfunction $\psi_0$ of $\Hag$ which
is positive almost everywhere. Such an eigenfunction is up to a
scalar multiple uniquely defined, i.e. the corresponding
eigenvalue is non-degenerate. We will show in
Lemma~\ref{l-boundgrst} that, if $\Hag$ is a Hamiltonian in
$\RR^2$ with a singular potential supported on a curve, such a
function $\psi_0$ exists and is the eigenfunction corresponding to
the lowest eigenvalue of $\Hag$. Let us define the unitary
transformation
\[
U \colon L^2 \rightarrow L^2_{\psi_0} := L^2(\RR^d ,
\psi_0^2\mathrm{d} x )\,,\quad Uf:=\psi_0^{-1}f\,,\quad f\in L^2
\]
and denote the eigenvalue corresponding to $\psi_0$ by $E_0$.
Furthermore, consider the sesquilinear form
\begin{equation} \label{e-unitaryform}
\tilde{\cE}_{\alpha \sigma_\Gamma}(\psi,\phi )= \cE_{\alpha
\sigma_\Gamma}(U^{-1}\psi,U^{-1} \phi )- E_0 (U^{-1}\psi,U^{-1} \phi )\,,
\end{equation}
for $\psi , \phi \in D(\tilde {\cE}_{\alpha \sigma_\Gamma}) =
W^{1,2}(\RR^d ,\psi_0 ^2\mathrm{d}x) $.

Similarly as for regular potentials, after the unitary
transformation the information about the singular potential is
comprised in the weighted measure, i.e. we have
\begin{thm}
\label{t-spspegap} The form $\tilde{\cE}_{\alpha \sigma_\Gamma}$ admits the
following representation
\begin{equation} \label{formrep}
\tilde{\cE}_{\alpha \sigma_\Gamma}(\psi,\phi ) =
\int_{\RR^d}(\nabla\psi )(\nabla \overline \phi )\, \psi_0^2
\mathrm{d}x\,,\quad \mathrm{for }\quad \psi , \phi \in D(\tilde
{\cE}_{\alpha \sigma_\Gamma})\,.
\end{equation}
\end{thm}
\begin{proof}
To show the claim let us consider first the form $\cE_{\alpha
\sigma_\Gamma}(U^{-1}\psi,U^{-1}\phi )$ for $\psi ,\phi\in C_0^ \infty
(\RR^d )$. Using (\ref{form}) we obtain by a straightforward
calculation
\begin{eqnarray}
&& \nonumber \cE_{\alpha \sigma_\Gamma}(U^{-1}\psi,U^{-1}\phi )= -\alpha
\int_{\RR^d}I_{\sigma_\Gamma} (\psi \overline \phi \psi_0^2 )
\mathrm{d}\sigma_\Gamma
+%\label{deriv1}
\\&&
\int_{\RR^d}\left[ (\nabla\psi )(\nabla \overline \phi)\psi_0^2
+(\nabla\psi )\overline \phi \psi_0\nabla \psi_0+ \psi(\nabla
\overline \phi) \psi_0\nabla \psi_0+\psi \overline \phi (\nabla
\psi_0)^2 \right] \mathrm{d}x\,.\label{deriv2}
\end{eqnarray}
The last term in the above expression can be expanded by
integrating by parts in the following way
\begin{eqnarray}
\label{bypa1} && \int_{\RR^d}\psi \overline \phi (\nabla \psi_0)^2
\mathrm{d}x =- \int_{\RR^d} \nabla (\psi \overline \phi \nabla
\psi_0 )\psi_0 \mathrm{d}x
\\
&& \nonumber + \int_{\RR^d} I_{\sigma_\Gamma } (\psi \overline \phi
\psi_0) (\partial_{\mathrm{n}}^-  \psi_0 + \partial_{\mathrm{n}}^+
\psi_0 )\mathrm{d}\sigma_\Gamma \,.
\end{eqnarray}
To deal with the last expression we expand by differentiation the
term of the r.h.s. on (\ref{bypa1}) onto three components, use
boundary conditions (\ref{e-boundcond}) and the fact that $\psi_0$
is the eigenfunction of the Laplacian with these boundary conditions.
Finally, putting together (\ref{deriv2}) and (\ref{bypa1}) and
inserting it to (\ref{e-unitaryform}) we obtain the equivalence
(\ref{formrep}) on $ C_0^ \infty (\RR^d )$. Extending it by
continuity to $D(\tilde {\cE}_{\alpha\sigma_\Gamma})$ we get the
claim.
\end{proof}

Since our aim is to estimate the spectral gap we assume that
\begin{equation}
\label{e-as2} \text{ the bottom of the spectrum of $\Hag$ consists
of two isolated eigenvalues.}
\end{equation}
Let $E_1= \inf_{\psi\perp \psi_0, \|\psi\|=1} \cE_{\alpha
\sigma_\Gamma}(\psi,\psi)$ denote the first excited eigenvalue and denote
by $\psi_1$ a corresponding eigenfunction. It follows from
\eqref{e-unitaryform} that
\begin{equation}
E_1 -E_0 = \tilde{\cE}_{\alpha \sigma_\Gamma}[U\psi_1 ]/ \|\psi_1\|^2\,,
\end{equation}
where we use the abbreviation $\tilde{\cE}_{\alpha \sigma_\Gamma}[\phi
]=\tilde{\cE}_{\alpha \sigma_\Gamma}(\phi,\phi )$. If $E_1$ is degenerate
the formula holds for \emph{any} eigenfunction. Using Theorem
\ref{t-spspegap}, we have
\begin{cor} \label{firstgap}
The spectral gap between the two lowest eigenvalues of $\Hag$ is
given by
\[
E_1 -E_0 = \int \l|\nabla \frac{\psi_1}{\psi_0}\r|^2
\psi_0^2\mathrm{d}x / \|\psi_1\|^2\,.
\]
\end{cor}
Let us note that notations $\psi_i$ correspond to
$\psi_{\kappa_i}$ where $E_{i}=-\kappa _i ^2$ from
Theorem~\ref{resolvent}.
\bigskip

\section{Estimates for the lowest spectral gap of Hamiltonians
with interaction on a finite curve} \label{s-curve}

The aim of this section is to derive explicit estimates for the
lowest spectral gap of $\Hag$. We will use the general results
obtained in Section \ref{s-general-gap}, and apply them to a two
dimensional system. More precisely, let $\Gamma \subset \RR^2$ be
a finite curve given as the range of the $C^2$-parameterization
$[0,L] \ni s\mapsto \gamma(s)=(\gamma_1 (s), \gamma_2 (s)) \in
\RR^2$ without self-intersections. (Exception: If $\Gamma$ is a
closed curve the starting and end point of the curve coincide. In
that case we also require that the first two derivatives of the
parameterization $\gamma$ coincide at the parameter values $0$ and
$L$.) We assume that $\Gamma$ is parameterized by arc length.

Denote by $\mG $ a constant satisfying $\mG <L$ if $\Gamma $ is
%\Hm{Maybe, instead of $m_\Gamma$ we can just write $m$ - this is
%just argument of $M_\gamma$; what do you think?}
closed and $\mG
\leq L$ otherwise. By the $C^2$-differentiability assumption on
$\gamma $, for each $\mG$, there exists a positive constant
$\Mg:=\Mg(\mG)$ such that
\begin{equation}\label{e-geoGamma}
\Mg(\mG) |s-s'|\leq |\gamma(s)-\gamma(s')|\quad \mathrm{for} \,\,\,
s\,,s'\in \RR \text{ with } |s-s'| \le \mG\,,
\end{equation}
where $|\gamma(s)| =\sqrt{\gamma_1 (s)^2 +\gamma_2 (s)^2}$.
We choose $\Mg(\mG)$ to be the largest possible number satisfying
inequality \eqref{e-geoGamma}. Then $\mG \mapsto \Mg(\mG)$ is a
non-increasing, continuous function.
\bigskip

The estimates we will derive depend on the geometry of the curve
through its length $L$, the diameter of $\Gamma$, the maximum
\[
K:= \max_{s \in [0,L]} \varkappa (s)
\]
of its curvature $\varkappa\colon [0,L] \to \RR$, and
the values $\Mg (L/2)$ and $M_{\tg} (L/2)$,
where $\tg$ is defined in \eqref{e-gammatilde}.

\begin{rem}
In fact our methods work also for $C^1$-curves which are piecewise
$C^2$ regular. This means that there are finitely many values
$0=:s_1, \dots,s_N:=L$ such that $\gamma\colon ]s_i, s_{i+1}[ \to
\RR^2$ is of class $C^2$ for each $i\in \{1, \dots,N-1\}$ and at
each point $\gamma(s_i)$ the curvature of $\Gamma$ jumps by an
angle $\varphi_i$. In this case the proofs become somewhat more
technical. The constants which under the global $C^2$-assumption
depend only on the curvature are in the more general case
additionally dependent on the angles $\varphi_2, \dots,
\varphi_{N-1}$.
\end{rem}

Following the general discussion given in Section~\ref{s-general}
we can construct the Hamiltonian $\Hag$ with a perturbation on
$\Gamma $ as the operator associated with the form $\cE_{\alpha
\sigma_\Gamma }$ given by (\ref{form}). Furthermore, the operator $\Hag$
is associated to the boundary conditions \eqref{e-boundcond}.

To derive estimates for the lowest spectral gap we will work in an
appropriate neighbourhood of $\Gamma$ and to this aim we will
introduce the following notation. For $\epsilon\ge0$ let
$\cC_\epsilon$ be a convex hull of the set $\Gamma_\epsilon:=\{
x\in \RR^2\mid \dist(x, \Gamma) \le \epsilon\}$. We denote
\[
\cC:= \cC_0 \,,\quad R := \inf\{r>0\mid \exists x \in \RR^2 :
B_r(x) \supset \cC_1 \}\,.
\]
Let $x_0\in \RR^2$ be such that $B_R :=B_R(x_0 ) \supset \cC_1$.

\begin{rem} \label{r-RandL}
For a connected curve $\Gamma$ we have clearly $R\le 1 + \frac{L}{2}$. In
the general situation, where $\Gamma$ consists of several
topological components this is no longer true, and $R$ and $L$ are
completely independent parameters of our model.
\end{rem}

We employ Corollary~\ref{firstgap} and the H\"older inequality to
obtain the lower bound for
\begin{equation} \label{1estim}
E_1 -E_0 \geq \frac{(\int_{B_R}|\nabla f
|\mathrm{d}x)^2}{\|\psi_0\|^2\|\psi_1\|^2} \inf_{x\in B_R }
\psi_0(x)^4\,,
\end{equation}
where $f:= \psi_1 /\psi_0$, cf.~\cite{KirschS-87}.

Set $\kappa_i :=\sqrt{-E_i}$ for $i=0,1$. The main result of this
section is contained in the following statement.

\begin{thm}\label{t-spgaps}
Suppose that assumption \eqref{e-as2} is satisfied. Then the
lowest spectral gap of $\Hag $ can be estimated as follows
\begin{equation}
E_1 - E_0 \ge \kappa^2_1\,\, \mu_{\Gamma ,\alpha }(\rho ,\kappa_0)
\,\mathrm e^{-C_0\rho}\,,\quad \rho :=\kappa_0 \, R
\end{equation}
where $\mu_{\Gamma ,\alpha }(\cdot,\cdot )$ is a polynomial
function and $C_0$ is an absolute constant.
\end{thm}

The precise formula \eqref{defmu} for the function
$\mu_{\Gamma,\alpha} $ is derived at the end of Section
\ref{s-gradient}. For the proof of this theorem we need several
 lemmata estimating the behavior of all ingredients
involved in the r.h.s. of (\ref{1estim}). They are collected in
the subsequent sections.

%%%%%%%%%%%%%%%%%%%%%%%%%%%%%%%%%%%%%%%%%%%%%%%%%%%
\section{Pointwise estimates on the eigenfunctions}
\label{s-pointwise}

\subsection{Lower bound for the ground state.}
The first step is to obtain a lower bound for $\inf_{x \in B_R}
\psi_0 (x)$. The sought estimate is given in the following

\begin{lem}
\label{l-boundgrst} \noindent (i) The ground state $\psi_0$ of
$\Hag $ is a simple eigenfunction.

\noindent (ii) The function $\psi_0$ is strictly positive on
$\RR^2$ and moreover we have
\begin{equation}
\label{estground} \inf_{x \in B_R} \psi_0 (x) \ge C_1 \kappa_0
\frac{\mathrm{ e}^{-2\rho }}{1+\sqrt{2\rho}}\| \psi_0 \|%_{L^1(\sigma_\Gamma)}
\,,\quad \mathrm{where } \quad \rho =\kappa_0 R\,,
\end{equation}
and $C_1$ is a positive constant.
%, $w_0$ is the function defined by
%(\ref{e-ef}), i.e. $\psi_0 =R_{\mathrm{d}x,\sigma_\Gamma }^{i\kappa_0}w_0$.
\end{lem}

\begin{rem} \label{rem-kernel}
It is useful to note that the integral kernel of the inverse of
the two dimensional Laplacian has the following representation
\begin{equation}\label{e-kernel}
G^{i\kappa }(x-x')=\frac{1}{(2\pi)^2}\int_{\RR^2}
\frac{\mathrm{e}^ {ip(x-x')}}{p^2+ \kappa^2 }\mathrm{d}p=
\frac{1}{2\pi} K_0 (\kappa |x-x'|)\,,
\end{equation}
where $K_0$ is the Macdonald function \cite{AbramowitzS-72}.
\end{rem}

\begin{proof}[Proof of Lemma \ref{l-boundgrst}]
\indent (i) To prove the theorem we will use the representation
$\psi_0 =R_{\mathrm{d}x,\sigma_\Gamma }^{\kappa_0}w_0$, cf.~(\ref{e-ef}).
Using the same argument as in \cite{Exner-05} we conclude that
$w_0$ is a simple, positive eigenfunction of $R^{\kappa
_0}_{\sigma_\Gamma ,\sigma_\Gamma}$, (the argument from \cite{Exner-05} can be
extended to curves which are not closed). This implies the
simplicity of $\psi_0$.

\noindent (ii) Since the kernel of $R^{\kappa_0}_{\mathrm{d}x
,\sigma_\Gamma}$ is a strictly positive function and $w_0$ is positive
%in view of Theorem~\ref{resolvent} (see parts (iii), (iv))
we have
\begin{equation} \label{estground1}
\psi_0 (x)=(R_{\mathrm{d}x,\sigma_\Gamma }^{\kappa_0}w_0)(x)=\int_{\RR^2}
G^{\kappa_0} (x-x' )w_0(x ')\mathrm{d}\sigma_\Gamma (x')>0\,.
\end{equation}
Furthermore, using the representation $G^{\kappa
}(\xi)=\frac{1}{2\pi} K_{0}(\kappa \xi)$ (see Remark
\ref{rem-kernel}) and the behavior of the Macdonald function
$K_0$, cf.~\cite{AbramowitzS-72}, one infers the existence of a
constant $C_2>0$ such that
\[
G^{\kappa }(\xi) >C_2\frac{\mathrm{e}^{-\kappa
\xi}}{1+\sqrt{\kappa\xi}}.
\]
Combining this with the positivity of $w_0$ and formula
(\ref{estground1}) we get
\[
\inf_{x\in B_R }\psi_0 (x)\geq \inf_{(x,x')\in B_R\times \Gamma }
G^ {\kappa_0 }(x-x')\|w_0\|_{L^1 (\sigma_\Gamma )} \geq
C_2(1+\sqrt{2\rho})^{-1} \mathrm{e}^{-2\rho}\|w_0\|_{L^1 (\sigma_\Gamma
)}\,.
\]
Moreover it follows from formula (\ref{e-norm2}) in Section
\ref{ss-norms} that $2^{3/2}\pi \kappa_0\|\psi_0\|\leq
\|w_0\|_{L^1 (\sigma_\Gamma )}$ which completes the proof.
\end{proof}

Let us note that the above result is analogous to the one obtained
for regular potentials in \cite{KirschS-87}. However the method
used there is mainly based on the Feynman--Kac formula which
cannot be directly applied to singular potentials.

\subsection{Localizations of zeros and maxima of eigenfunctions}
To obtain an estimate on the gradient which is involved in
(\ref{1estim}) we need some informations on the behaviour of the
functions $\psi_0$ and $\psi_1$. For this aim we will localize
their zeros and maxima.

Let us recall that $v$ is a \emph{subsolution}, respectively
\emph{supersolution}, of the equation $(-\Delta-E)u = 0$ in an
open set $\Omega$, if $ (-\Delta-E)v(x) \le 0$, respectively
$(-\Delta-E)v(x) \ge 0$, for all $x \in \Omega$. In the sequel we
need the following fact, see e.g.~Lemma 2.9 in \cite{Agmon-85}.
\begin{lem}
\label{l-subsolution+} Let $v$ be a subsolution of the equation
$(-\Delta-E)u = 0$ in an open set $\Omega$. Then $v_+:= \max\{v,
0\}$ is also a subsolution of the same equation in $\Omega$.
\end{lem}

Our next task is to localize the maxima, minima and zeros of
eigenfunctions of $\Hag$.

\begin{prp}\label{p-maximazeros}
Let $\psi$ be a real eigenfunction of $\Hag$ with negative
eigenvalue $E$. Then all its maxima and minima lie on $\Gamma$. If
$\psi$ is not the ground state, at least one zero of $\psi$ lies
in $\cC$.
\end{prp}
The analog of the proposition holds for proper potentials as well
as singular ones in arbitrary space dimension.

\begin{proof}
Let $\psi$ be any eigenfunction of the operator $\Hag$ to the
eigenvalue $E<0$. Then for any $\epsilon >0$, on the complement of
$\Gamma_\epsilon$ we have $-\Delta \psi = E \psi$. For $\psi_+ :=
\max\{\psi, 0\}, \psi_- := \max\{-\psi, 0\}$ we have again
\begin{equation}
- \Delta \psi_+ \le E \psi_+ \quad - \Delta \psi_- \le E \psi_-
\quad\text{ on } \RR^2 \setminus \Gamma_\epsilon\,.
\end{equation}
By the strong maximum principle, $\psi_+, \psi_-$ assume their
maxima inside $\Gamma_\epsilon$, unless they are constant, cf.~for
instance Thm.~2.2 in \cite{GilbargT-83}. The latter can only occur
if we consider a bounded region with boundary conditions which are
different from Dirichlet ones. Thus the minima and maxima of
$\psi$ are contained in $\bigcap_{\epsilon>0}
\Gamma_\epsilon=\Gamma$. If $\psi $ is not the ground state, at
least one of its zeros is contained in $\cC$ since $\psi$ is real.
\end{proof}

\subsection{Relation between norms}\label{ss-norms}
\begin{lem}\label{lem-norms}
Let $ \psi$ be an eigenfunction of $\Hag$ to $E=-\kappa^2$. We
have
\begin{equation}\label{e-norm1}
\|\psi\|\leq \ \frac{L\alpha }{2^{3/2}\pi \, \kappa }\|\psi
\|_{\infty } \,,
\end{equation}
where $L$ is the length of $\Gamma$.
\end{lem}
\begin{proof}
To prove the claim we will use the representation
\[
\psi (x)\equiv \psi_\kappa (x)=(R_{\mathrm{d}x,\sigma_\Gamma
}^{\kappa}w _\kappa)(x)=\int_{\RR^2} G^{\kappa} (x-x' )w _\kappa
(x ')\mathrm{d}\sigma_\Gamma (x')\,,
\]
cf.~(\ref{estground1}) and the Fourier transform of $G^{\kappa}$
given by (\ref{e-kernel}). A straightforward calculation yields
\begin{equation}\label{e-norm2}
\|\psi\|^2\leq \frac{1}{(2\pi )^3}\int_{\RR^2} \frac{1}{(p^2
+\kappa^2)^{2}}\dd p \|w\|^2_{L^1 (\sigma_\Gamma )} =\frac{1}{2^3
\pi^2}\kappa^{-2}\|w\|^2_{L^1 (\sigma_\Gamma )}\,.
\end{equation}
Combining relation (\ref{e-relationpsiw}) with the fact that the
maxima and minima of $\psi$ lie on $\Gamma $ we obtain
\begin{equation}\label{e-norm3}
\|w\|_{L^1 (\sigma_\Gamma )}\leq L \|w\|_{\infty }=\alpha
L\|\psi\|_{\infty}\,.
\end{equation}
Applying (\ref{e-norm3}) to (\ref{e-norm2}) we get the desired
inequality.
\end{proof}
\bigskip

\begin{lem}\label{lem-estnormpsi}
Let $\psi$ be an eigenfunction of $\Hag $ with the corresponding
eigenvalue $-\kappa^2$. There exists a positive absolute constant
$\eta_0 $ such that
\begin{equation}\label{e-norm4}
\|\psi \| \geq \frac{\kappa R}{(\cG_5\alpha
+1)^2(\kappa^2+1)}\mathrm{e}^{-\eta_0 \kappa R} \|\psi\|_\infty\,,
\end{equation}
where $\cG_5:=\max\{2\cG_4,\cG_3+\cG_4 \log(\max\{1,L\})\} $ and
$\cG_3$ and $\cG_4$ are taken from Corollary~\ref{c-logdis}.
\end{lem}
\begin{proof}
In the sequel we will use the following fact: If the curve
$\Gamma$ is parameterized by arc length, then at each point $x\in
\Gamma$ the vector tangential to $\Gamma$ and the unit normal
vector  form an orthogonal basis.

Denote by $v$ a point in $\RR^2$ where $\psi$ assumes its 
maximal value, i.e.~$\psi(v) = \|\psi\|_\infty$. We know from Lemma
\ref{p-maximazeros} that $v\in\Gamma$. Denote by $A$ the square
centered at $v$ with sidelength $2b$ whose sides are parallel to
the tangential, respectively the normal vector of the curve
$\Gamma$ at the point $v$. Let $x$ be an arbitrary point in $A$.
To estimate $|\psi(x)-\psi(v)|$ we would like to apply the
fundamental theorem of calculus to the gradient of $\psi$ and an
auxiliary curve connecting $x$ and $v$. The natural choice would
be a line segment joining the two points, however this segment
might be tangential to $\Gamma$. In this geometric situation we do
not have good control of $\nabla \psi$.

For this reason we consider the following picewise linear curve
connecting $v$ and $x$: join $x$ by a linear segment parallel to
the normal vector of $\Gamma$ at the point $v$ to the boundary of
$A$, do the same for $v$. Along this boundary edge of $A$ joint
the two line segments by a third line segment. This curve has at
most length $4b$

Now we complete the proof of the lemma along the lines of the
proof of Proposition \ref{p-intgrad} using the upper bound on the
gradient obtained in Proposition \ref{p-estgrad}. In the present
situation the argument is actually somewhat simpler than in the
proof of Proposition \ref{p-intgrad}. One has to choose the
parameter $b=\rho \l((\cG_5\alpha
+1)^2(\kappa^2+1)\r)^{-1}\mathrm{e}^{-\eta_0 \rho}$ with an
appropriate positive constant $\eta_0$, cf.(\ref{e-difference}),
and then establish analogues of the inequalities
\eqref{e-difference} and \eqref{e-1/4}.

This argument implies that
\[
|\psi(x)-\psi(v)| \le \frac{1}{2} \|\psi\|_\infty
\]
for all $x \in A$ if the sidelength of $A$ obeys
\[
2b = 2 \frac{\kappa R
}{(\cG_5\alpha+1)^2(\kappa^2+1)}\mathrm{e}^{-\eta_0 \rho}\,.
\]
Finally, by a straightforward calculation we get
\[
\|\psi\|^2 \geq \int_A \psi (x)^2\dd x\geq b^2 \|\psi\|_\infty^2
\,,
\]
which completes the desired result.
\end{proof}

\subsection{Exponential decay of eigenfunctions}

The following result on the exponential decay of eigenfunctions
from \cite{Agmon-85} will be useful in the sequel.

\begin{lem}\label{decaypsi}
Let $\tilde R >R$ and $\psi$ be an eigenfunction of $\Hag$
corresponding to the eigenvalue $E=-\kappa^2$, where $\kappa >0$.
Set $\phi(x) = \phi(|x|)= \sqrt{{\tilde R}/|x|} \ e^{-\kappa (|x|-
\tilde R)}$. Then the following estimate holds
\[
|\psi(x)| \le \|\psi\|_\infty \phi(x) \,\,\,\,\mathrm{ for
}\,\,\,\, x \in D^c :=\RR^d \setminus D, \,\,D:= \{x \mid |x| <
\tilde R \}\,.
\]
\end{lem}

\begin{proof}
$\phi$ is a supersolution and $|\psi|$ is by Lemma
\ref{l-subsolution+} a subsolution of the equation $(-\Delta -E)
u=0$ in $D^c$. Thus for any constant $C_4 > 0$ the function
\begin{equation}
F= (C_4 \phi - |\psi|)_-
\end{equation}
is a supersolution. We can choose the constant
$C_4\ge \sup_{|x|=\tilde R}|\psi(x)|$ such that $F$ vanishes
identically on $\partial D$. The maximum principle implies
that the supremum of $ F$ on the closed set
$D^c$ is assumed at its boundary. Therefore
\[
F\equiv 0 \text{ on } D^c\,.
\]
This implies the statement of the lemma.
\end{proof}

\section{Estimates on the gradient of eigenfunctions}
\label{s-gradient}

The aim of this section is to derive a lower bound for the
expression $\int_{B_R }|\nabla f|\dd x$ involved in (\ref{1estim})
where $f:=\psi_1/\psi_0$.

The strategy here is the following. First we will derive upper
bounds for $\nabla \psi_0$, $\nabla \psi_1$. Combining this with
the inequality
\begin{equation} \label{e-leqgrad}
|\nabla f | \le \frac{|\psi_1| \ |\nabla \psi_0|}{\psi_0^2} +
\frac{|\nabla\psi_1|}{\psi_0}\,
\end{equation}
and using Lemma~\ref{l-boundgrst} we will get an upper bound for
$\nabla f$. The estimate on the gradient gives a quantitive upper
bound for the variation of the function $f$ and is used in
Proposition \ref{p-intgrad} to provide a lower bound for
$\int_{B_R }|\nabla f|\dd x$.
\subsection{Preliminary estimates on certain integrals}
To derive upper and lower bounds on gradients the following
geometric notions and generalized distance functions will prove
useful.

\begin{dfn} \label{d-admissible}
Let $S$ be a line segment of the length $2b$ intersecting the curve
$\Gamma$ at the mid point.  We call $y \in
\RR^2\setminus \Gamma$ an \emph{$S$-admissible point}, if the
following hold:
\\
there is a unit vector $e$ parallel to $S$ (up to orientation)
such that
\[
\Theta_S := \{ y-t e \mid t \in [0, \infty[\}
\]
intersects $\Gamma$. Denote $t_1:= \min \{ t \in[0,\infty[\mid
y-te\in \Gamma\}$ and let $s_y\in[0,L]$ be the parameter value
such that
\[
\gamma(s_y) = y-t_1 e  \quad (\in \Gamma \cap \Theta_S) \, .
\]
Denote by $\theta$ the angle at which $\Theta_S$ and $\Gamma$
intersect at $\gamma(s_y)$, more precisely
\[
\cos \theta := \la e, t(s_y)\ra
\]
where $t(s_y) := \gamma'(s_y)\in \RR^2$ is the (unit) tangential
vector to the curve $\Gamma$ at the point $\gamma(s_y)$. Assume
that the angle $\theta$ is neither zero nor $\pi$, i.e.
\[
|\cos \theta| < 1 \, .
\]
Denote by $d_S(y) = t_1$ the distance between $y$ and
$\gamma(s_y)$, which is also the distance form $y$ to $\Gamma$
along $\Theta_S$. Thus for any $y \in S$ we have $d_S(y) \le b$.
\end{dfn}
\begin{rem} \label{r-pi/6}
In our application in the proof of Proposition \ref{p-intgrad}. we
will only need to consider line segments which intersect $\Gamma$
at an angle which is at least $\pi/6$, i.e.~we have $|\cos
\theta|\le \sqrt{3}/2$. Therefore we assume this bound in the
sequel.
\end{rem}

In the following we assume that $S$ is a line segment intersecting
$\Gamma$ and $ y \in \RR^2\setminus\Gamma$ an $S$-admissible point
with vector $e$ in the sense of Definition \ref{d-admissible}.
%Moreover we use the notation $ \Mg (\cdot)$ for the function
%defined in Section~\ref{s-curve}.

\begin{lem}  \label{l-TaylorCosine}
Define the function $\phi\colon [0,L] \to \RR$ as the angle
$\phi(s)$ between the vector $e$ and the vector
$\gamma(s)-\gamma(s_y)$, more precisely
\[
g(s):=\cos \phi(s) := \la e, \tilde t(s) \ra\,,
\]
where $\tilde t(s):= \tau(s) / | \tau(s)|$ is the normalization of
the vector $\tau(s) := \gamma(s)-\gamma(s_y)$. Then we have
\begin{equation} \label{e-gApprox}
|g(s) - g(s_y) | \le \frac{2 K}{\Mg (\mG )} |s-s_y| \quad \mathrm{
for } \quad |s-s_y|  \leq \mG\,.
%\leL/2\,, where
\end{equation}
\end{lem}
%Here we have chosen $\mG=L/2$ in equation \eqref{e-geoGamma},
%in other words $\Mg=\Mg(\mG)=\Mg(L/2)$. A more
%careful analysis using the arc-length parameterization of the
%curve would eliminate the $\Mg$ dependence in the estimate
%\eqref{e-gApprox}.

\begin{proof}
In the proof we will denote by $s'$ a generic value in the
interval $[s_y, s]$ and it may change from estimate to estimate.

First we calculate and derive a bound for $g'$. A calculation
using the product rule gives
\[
\frac{d}{ds} \Big ( \frac{\tau (s)}{|\tau (s) |} \Big ) =
\frac{\tau'(s) \langle \tau(s), \tau(s)\rangle - \tau  \langle
\tau(s), \tau'(s)\rangle} {(\langle \tau(s),
\tau(s)\rangle)^{3/2}}\,.
\]
This can be expressed more geometrically by means of the
orthogonal projection $P(s)\colon \RR^2\mapsto\RR^2$ onto the line
orthogonal to the vector $\tau(s)$. The formula is $P(s)=I -
\frac{|\tau\ra \la \tau |}{\la \tau(s), \tau(s)\ra} = | \tilde
n(s) \ra \la \tilde n(s)| $, where we use the Dirac brac-ket
notation, $\tilde n(s)$ denotes a unit vector perpendicular to
$\tilde t(s)$ and $I$ is the identity operator. Then we have
\[
\frac{d}{ds} ( \tilde t(s) ) = \frac{P(s) \tau'(s)} {|\tau(s)
|}\,.
\]
Now, since $\tau(s) = \gamma(s)-\gamma(s_y)$ we obtain $\tau'(s) =
\gamma'(s)$ and consequently
\[
g'(s) =\frac{\langle e, \tilde n(s)\ra \la \tilde n(s),
\gamma'(s)\rangle }{ |\tau(s) |}\,.
\]
For $s$ close to $s_y$ we expect $\la \tilde n(s),
\gamma'(s)\rangle \approx \la n(s_y), \gamma'(s_y)\ra=0$, where
$n(s)$ is a unit normal vector to the curve  at the point
$\gamma(s)$. Let us make this more precise.

From Taylor's formula for the curve $\gamma\colon [0,L] \to \RR^2$
it follows
\begin{gather*}
\tau(s) = \gamma(s) -\gamma(s_y) = (s-s_y) t(s_y) +
\frac{(s-s_y)^2}{2} \gamma''(s'), \text{ for } s'\in [s_y,s]
\end{gather*}
and therefore we get
\begin{gather*}
\Big | \frac{\tau(s)}{s-s_y}- t(s_y) \Big | \le \frac{K}{2} |
s-s_y|\,.
\end{gather*}
This gives the following estimates
\begin{gather*}
\big||\tau(s)|  - |s-s_y|\big| \le |\tau(s)  - (s-s_y) t(s_y) |
\le \frac{(s-s_y)^2}{2} |\gamma''(s')| \le \frac{K \,
(s-s_y)^2}{2}
\end{gather*}
and
\begin{gather*}
\Big | \frac{\tau(s)}{s-s_y} - \frac{\tau(s)}{|\tau(s)|}\Big | \le
|\tau(s)|  \Big | \frac{|\tau(s)|-(s-s_y)}{(s-s_y)\tau(s)}\Big |
\le \frac{K \, |s-s_y|}{2}\,,
\end{gather*}
which, in turn, imply
\[
| \tilde t(s) - t(s_y)| \le  \Big | \frac{\tau(s)}{ |\tau(s) |}-
\frac{\tau(s)}{s-s_y}\Big | +\Big | \frac{\tau(s)}{s-s_y}  -
t(s_y)\Big | \le K |s-s_y|\,.
\]
An easy calculation shows that $|\tilde n (s) - n(s_y)| =|\tilde
t(s) - t(s_y)|$. Combining this we the above inequalities we can
estimate the sought expression for $ |s-s_y| \le \mG$
\begin{multline*}
|g'(s)| = \frac{1}{\
 |\tau(s) |} \, |\langle e, \tilde n(s)\ra \la \tilde n(s), \gamma'(s)\rangle|
\le \frac{1}{\Mg (\mG)|s-s_y|} \, |\la \tilde n(s),
\gamma'(s)\rangle|
\\
\le \frac{1}{\Mg (\mG)|s-s_y|} \, \Big( |\la n(s_y),
\gamma'(s)\rangle|  + | n(s_y) - \tilde n(s) |\,   |\gamma'(s) |
\Big)\,.
\end{multline*}
Using the formula $ \gamma'(s)= \gamma'(s_y) + (s-s_y)
\gamma''(s')$ for $s' \in [s_y,s]$ we arrive at
\begin{align*}
|g'(s)| \le \frac{2 K}{\Mg (\mG)} \,.
\end{align*}

Since $g$ is continuously differentiable, by Taylor's formula
there is a number $s'\in [s_y, s]$ such that
\[
g(s) = g(s_y) + (s-s_y) g'(s')\,.
\]
This finally implies
\[
|g(s) - g(s_y)| \le  |s-s_y| \| g'\|_\infty  \le \frac{2
K}{\Mg (\mG )} |s-s_y|\,.
\]
\end{proof}

\begin{lem}  \label{l-TaylorDist}
Let $S$ be a line segment intersecting $\Gamma$ and $ y \in
\RR^2\setminus\Gamma$ an $S$-admissible point with vector $e$ in
the sense of Definition \ref{d-admissible}. Set
$\delta_0=\delta_0(\theta,K, \Mg (L/2),L)=
\min\{\frac{L}{2},\frac{\Mg(L/2)}{2K}\frac{1-|\cos\theta|}{2}\}$
and $\tau \ge \frac{1}{2} (1-|\cos \theta|)>0$.
Then we have for all $s \in
[s_y-\delta_0,s_y+\delta_0] \cap [0,L]$
\[
|y-\gamma(s)|^2
\ge \tau\ (d_S(y)^2 + |\gamma(s_y)-\gamma(s)|^2)\,.
\]
\end{lem}

\begin{proof}
Since
\[
\tilde t(s)=\frac{\tau(s)}{|\tau(s)|} =
\frac{\gamma(s)-\gamma(s_y)}{|s-s_y|} \,
\Big|\frac{s-s_y}{\gamma(s)-\gamma(s_y)}\Big|\,,
\]
the equalities $\lim_{s \to s_y} \tilde t(s)= \tau (s_y) \, |\tau
(s_y)|^{-1}= t(s_y)$ and $\lim_{s \to s_y} g(s) = \lim_{s \to s_y}
\la e, \tilde t(s) \ra=\la e, t(s_y) \ra= \cos\theta$ hold.
For $\delta_0$ as in the statement of the lemma we have
by \eqref{e-gApprox}
\[
|g(s) - \cos \theta| \le  \frac{1}{2}(1- |\cos \theta|) \quad
\text{for all}\quad |s-s_y| \le \delta_0\,,
\]
which implies $|g(s)| \le \frac{1}{2}(1+ |\cos \theta|)<1$. Now
the cosine formula gives us
\[
|y -\gamma(s)|^2 = d_S(y)^2 + |\gamma(s_y)-\gamma(s)|^2 - 2 d_S(y)
|\gamma(s_y)-\gamma(s)| g(s)\,.
\]
Set now $\tilde \tau := 1- |g(s)|$. By the definition of $\delta_0$
we have $\tilde \tau \ge \tau$, which is positive
since $|\cos \theta|<1$ by Definition \ref{d-admissible}.
Therefore the binomial formula implies
\[
|y -\gamma(s)|^2 \ge \tilde \tau \l( d_S(y)^2 +
|\gamma(s_y)-\gamma(s)|^2 \r)\,.
\]
\end{proof}

\begin{rem}\label{r-gammatilde}
In the following we will need a lower bound for
$|(\gamma(s )-\gamma(s_y)(s-s_y)^{-1}|$ uniform with respect to
$s\in [0,L]$. Such a lower bound does not exists if $\Gamma$ is a
closed curve parameterized by $\gamma$ in such a way that $s_y =0$.
In this case define a new parameterization $\tg\colon [0,L] \to \RR^2$ by
\begin{equation}
\label{e-gammatilde}
\tg(s) := \begin{cases}
      \gamma (s + \frac{L}{2}) \text{  for } s \in  [0,\frac{L}{2}[\\
      \gamma (s - \frac{L}{2}) \text{  for } s \in  [\frac{L}{2},L].
          \end{cases}
\end{equation}
It is easily seen that any arc segment $\hat\Gamma \subset \Gamma$ of length $L/2$ or less
which contains the point $\gamma(0)$ in its interior (relative to the set $\Gamma$)
cannot contain the point $\gamma(L/2)=\tg(0)$. This shows that
\begin{equation}
\label{e-gammamajorize}
\inf_{\gamma_1} M_{\gamma_1}(\mG)
%=\min_{\gamma_2\in \{\gamma,\tg\}} M_{\gamma_2}(\mG)
= \min\{\Mg (\mG), M_{\tg} (\mG) \}
=:\tilde M(\mG) \quad \text{
for } m_\Gamma \le  L/2 \,,
\end{equation}
where $\gamma_1$  runs over all arc length parameterizations of $\Gamma$.
Thus $M:=\tilde M(L/2)$ can be used for all
parameterizations as an lower bound in \eqref{e-geoGamma} for intervals no longer then $L/2$.

Choose now
a parameterization $\gamma_1$ of the curve such that $s_y =L/2$.
This implies that for any $s \in [0,L]$ we have $|s-s_y|\le L/2$
and thus $|\gamma_1 (s )-\gamma_1 (s_y)|\geq M_{\gamma_1}(L/2)|s-s_y|
\geq M|s-s_y|$.
\smallskip

In the following we will again write $\gamma$ instead of
$\gamma_1$, but the subsequent estimates are not affected
by this change since we have the universal bound \eqref{e-gammamajorize}.
%increased the constant
%$\Mg(\mG)$, $\mG \le L/2$ in a
%controlled way by changing the parameterization,
%cf.~\eqref{e-gammatilde} and \eqref{e-gammamajorize}. Similarly,
For the same reason we will also suppress the dependence on the
parameterization of some constants $\cG_i$ which depend on
the value $\Mg(L/2)$, since it can be bounded independently of the
chosen parameterization $\gamma$ using $M$.
\smallskip

In the next lemma we will use the abbreviation
$\cA_{\delta_0}:=[s_y- \delta_0, s_y+\delta_0]\cap [0,L]$,
$A_{\delta_0}:= \gamma(\cA_{\delta_0})$.
\end{rem}

%The following estimate will be useful in the proof of the next
%lemma. Let $y\in B_R\setminus\Gamma$ be $S$-admissible, then
%\begin{equation}
%\label{e-dSy-bound} d_S(y)^2 \le 10 R^2 \,.
%\end{equation}
%To see this use the cosine and the binomial formula to bound
%\begin{eqnarray*}
%d_S(y)^2 = |y-\gamma(s)|^2 + |\gamma(s_y)-\gamma(s)|^2 - 2 |y
%-\gamma(s)| |\gamma(s_y)-\gamma(s)| \cos \varpi
%\\
%\le 2 \l( |y -\gamma(s)|^2 + |\gamma(s_y)-\gamma(s)|^2 \r)\,,
%\end{eqnarray*}
%where $\varpi$ is the suitable angle between the segments
%$\overline{\gamma(s_y)\, \gamma(s)}$ and $\overline{y\,
%\gamma(s)}$. Obviously $|\gamma(s_y)-\gamma(s)| \le 2R$ and, if
%$\gamma(s)$ is the closest point to $y$ on $\Gamma$, we have $|y
%-\gamma(s)| \le R$. Thus \eqref{e-dSy-bound} is proven.

\begin{lem}
\label{l-logdis} (i) Let $y \in B_R \setminus \Gamma$ be an
$S$-admissible point. For $M$ as in Remark \ref{r-gammatilde}
and $|y-\gamma(s_y)| \le b_1:= M\delta_0/2$ we have
\begin{equation}
\frac{1}{2\pi}\int_{A_{\delta_0}}\frac{\dd \sigma_\Gamma(x)}{|y-x|} \le \cG_1
- \cG_2 \log d_S(y)\,,
\end{equation}
where the constant $\cG_1$ depends only on $\theta, K, M, L$
and $\cG_2$ only on $M$.

\noindent (ii) Moreover we have
\begin{equation} \frac{1}{2\pi}\int_{\Gamma\setminus
A_{\delta_0}}\frac{\dd \sigma_\Gamma(x)}{|y-x|} \le \frac{2}{\pi M } (\log L +
|\log \delta_0 |)\,.
\end{equation}
\end{lem}
If $\Gamma$ is not a closed curve, we can replace $M$ by $\Mg(L)$
by the monotonicity of the function $\Mg(\cdot)$.
\begin{proof}
We first prove statement (i).
Since $|s_y-s|\leq \delta _0\le L/2$ we get $|\gamma(s_y)-\gamma
(s)| \ge M |s_y-s|$.
Using now Lemma \ref{l-TaylorDist} we obtain
\[
|y-\gamma(s)|^2 \ge \tau \l (d_S(y)^2 + M^2 |s_y-s|^2 \r)\,.
\]
Applying the above inequality we can estimate the integral
\begin{multline*}
\frac{1}{2\pi} \int_{A_{\delta_0}} \frac{\dd \sigma_\Gamma(x)}{|y-x|}
=
\frac{1}{2\pi}\int_{\cA_{\delta_{0} }} \frac{\dd s}{|y-\gamma(s)|}
\le
\frac{1} {\pi M \sqrt{\tau }}
\int_{0}^{\delta_0} \frac{\dd \tilde s}{ \sqrt{(d_S(y) / M)^2+\tilde s^2}}
\\
= \frac{1}{\pi M\sqrt{\tau}} \l( - \log (d_S(y) / M) + \log
\l(\delta_0  +\sqrt{(d_S(y) / M)^2 + \delta_0 ^2} \r) \r)
\\
\le \frac{1}{\pi M\sqrt{\tau }} \l( - \log d_S(y) +
 \log M + \log \l( \l( 1+\frac{\sqrt{5}}{2}\r) \delta_0\r)
\r)\,,
\end{multline*}
where in the last estimate we use the inequality $d_S(y)\leq b_1$.
Using the explicite formula for $\delta _0$
one can check that the dependence of the constants
$c^\Gamma _1\,,c^\Gamma _2$ is precisely as stated in the lemma.
\smallskip

(ii) First, let us note that for all $s \in
[0,L]\setminus \cA_{\delta_{0}}$ and $|y-\gamma(s_y)| \le
b_1$ we have
\[
|y -\gamma(s)| \ge |\gamma(s_y) -\gamma(s)| -|y -\gamma(s_y)| \ge
M |s_y-s| -\frac{M \delta_0}{2} \ge \frac{M|s_y-s|}{2}\,.
\]
Consequently we get the following inequality
\begin{align*}
\frac{1}{2\pi} \int_{\Gamma\setminus A_{\delta_0}} \frac{\dd \sigma_\Gamma
(x)}{|y -\gamma(s)|} \le \frac{1}{\pi M} \int_{[0,L]\setminus
\cA_{\delta_{0}}} \frac{\dd s}{|s_y -s|}
\\
\le \frac{2}{\pi M} \int_{\delta_0}^L \frac{\dd s}{s} \le
\frac{2}{\pi M} (\log L + |\log \delta_0|)\,,
\end{align*}
which completes the proof.
\end{proof}

\begin{cor} \label{c-logdis}
Let $ y \in \RR^2\setminus\Gamma$ be an $S$-admissible point in
the sense of Definition \ref{d-admissible}. Assume that $d_S(y)
\le \min\{1,b_1\}$ and $|\cos \theta |\le \sqrt{3}/2$.
 Then there exist constants $\cG_3
=\cG_3(M,K)$ and $\cG_4 =\cG_4(M)$ such that
\begin{equation}
\frac{1}{2\pi}\int_{\Gamma}\frac{\dd \sigma_\Gamma(x)}{|y-x|} \le \cG_3 +
\cG_4 (\log L +|\log d_S(y)|)\,.
\end{equation}
\end{cor}

\subsection{Upper bound on the gradient}
In the sequel we will be interested in the behavior of the
gradient of an eigenfunction in some neighbourhood of $\Gamma $.
To this end we consider a line segment $S$ intersecting $\Gamma$
and an $S$-admissible point $y\in \RR^2\setminus \Gamma$
as in Definition \ref{d-admissible}.
We assume  $b\leq \min\{1,b_1\}$ and $|\cos \theta |\le \sqrt{3}/2$.
\begin{prp} \label{p-estgrad}
Let $\psi$ be an eigenfunction of $\Hag$ corresponding to the
negative eigenvalue $E=-\kappa^2 $. Let $S$ be a line segment
intersecting $\Gamma$ and $y\in \RR^2\setminus \Gamma$ an
$S$-admissible point. Then we have with the notation from
Definition~\ref{d-admissible} and Lemma~\ref{l-logdis} (ii): if $b
\le \min\{1,b_1\}$, then
\begin{equation}
\label{e-estgrad1} |\nabla \psi(y)| \leq \left [\cS_{\kappa ,
\Gamma }+\alpha \cG_4 |\log d_S(y)| \right]\|\psi\|_\infty \,,
\end{equation}
where
\[
\cS_{\kappa ,\Gamma }:=2\kappa^2(R+1)+C_5(\kappa R)^{1/2}
\mathrm{e}^{\kappa R}+\alpha \l(\cG_3+\cG_4 \log(\max\{1, L\})\r)
\,,
\]
$\cG_3$ depends only on $ M,K$ and $\cG_4$ depends only
on $M$.
\end{prp}

\begin{rem}
For the reader's convenience let us write down the estimate
\eqref{e-estgrad1} in the case that $S$ is perpendicular on
$\Gamma$, $\theta = \pi/2$ and $d_S(y)=\dist(y,\Gamma)$. Then we
have
\[
|\nabla \psi(y)| \leq \left [ \cS_{\kappa ,\Gamma }+\alpha \cG_4
|\log \dist(y,\Gamma)| \right]\|\psi\|_\infty \,.
\]
\end{rem}
\begin{proof}
The fundamental solution of the Laplace equation in two dimensions
is given by
\[
\fs(x,y) = \fs(|x-y|) = \frac{\log|x-y|}{2 \pi}\,.
\]
Let $\Omega$ be a domain in $\RR^2$ and $u \in C^2 (\bar \Omega)$.
Then the Green's representation formula
\[
u(y) = \int_{\partial \Omega} \left ( u(x) \frac{\partial
\fs(x,y)}{\partial \nu_x} - \fs(x,y) \frac{\partial u(x)}{\partial
\nu_x} \right)\dd \sigma_\Gamma (x) + \int_\Omega \fs(x,y) \Delta
u(x)\dd x \text{ for } y \in \Omega
\]
holds. Here $\mathrm{d}s(x)$ denotes the surface element and
$\frac{\partial}{\partial \nu_x} $ the outer normal derivative at
$x$. Let $\Omega$ be a bounded domain with positive distance to
the curve $\Gamma$. Consequently, for the eigenfunction $\psi$
satisfying $\Delta \psi=\kappa^2\psi$ on $\Omega $ the Green's
representation formula implies
\[
\psi (y) = \kappa^2 \int_\Omega \fs(x,y) \psi(x)\mathrm{d}x +
\int_{\partial \Omega} \left ( \psi(x) \frac{\partial
\fs(x,y)}{\partial \nu_x} - \fs(x,y) \frac{\partial
\psi(x)}{\partial \nu_x} \right)\mathrm{d}\sigma_\Gamma (x)\,.
\]
Now choose a monotone increasing sequence $\Omega_n ,n\in \NN$ of
domains as above such that $\bigcup_n \Omega_n = \Gamma^c$, where
$\Gamma^c$ stands for the complement of $\Gamma$. Then we have for
any $y \in \Gamma^c$
\begin{multline}
\label{e-reprpsi}
 \psi(y) =
\\= \lim_{n \to\infty}\left (
\kappa^2\int_{\Omega_n} \fs(x,y) \psi(x)\dd x + \int_{\partial
\Omega_n} (\psi(x) \frac{\partial\fs(x,y)}{\partial\nu_x} -
\fs(x,y) \frac{\partial \psi(x)}{\partial \nu_x})\mathrm{d}
\sigma_\Gamma(x) \right )
\\ 
 = \kappa^2 \int_{\RR^2} \fs(x,y) \psi(x) \dd x- \alpha
\int_\Gamma \fs(x,y) \psi(x)\mathrm{d} \sigma_\Gamma(x)\,.
\end{multline}

Here we have used several facts. Firstly, given $y\in \Gamma^c$
the functions $\partial_{\nu_x}\fs (\cdot ,y)$, $\fs (\cdot ,y)$
and $\psi(\cdot)$ are continuous. Secondly, the part of the
boundary $\partial \Omega_n$ which tends to infinity has a
vanishing contribution to the integral in the limit $n \to\infty$.
The remainder of the boundary $\partial \Omega_n$ tends for $n
\to\infty$ to two copies of $\Gamma$ with opposite orientation,
i.e.~opposite outward normal derivative, and formula
\eqref{e-boundcond} holds. In view of the exponential decay
established in Lemma~\ref{decaypsi} the first term in the last
line of (\ref{e-reprpsi}) is finite. Now, taking the gradient of
$\psi$ and using the chain rules we obtain
\begin{equation}
\nabla \psi(y) = \kappa^2 \int_{\RR^2} \nabla_y \fs(x,y) \psi(x)
\mathrm{d}x- \alpha \int_\Gamma \nabla_y \fs(x,y)
\psi(x)\mathrm{d}\sigma_\Gamma(x)\,.
\end{equation}
To deal with the singularity $\nabla_y \fs(x,y) =
\frac{1}{2\pi|x-y|}$ we split the integral over $\RR^2$ in two
regions, the ball $B_{R+1} =B_{R+1}(x_0)$ and its complement
$B_{R+1}^c$. Employing Lemma~\ref{decaypsi} we get
\begin{multline} %\label{e-estgrad}
|\nabla \psi(y)| \le \kappa^2 \left( \int_{B_{R+1}} \frac{1}{2\pi
|x-y|}\dd x + \int_{B_{R+1}^c} \frac{1}{2\pi |x-y|} \phi(x) \dd x
\right )\|\psi\|_\infty
\\
+ \int_\Gamma \frac{\alpha}{2\pi |x-y|}\dd
\sigma_\Gamma(x)\|\psi\|_\infty\,.
\end{multline}
The first integral can be estimated by
\[
\int_{B_{R+1}} \frac{1}{2\pi |x-y|}\dd x \leq
\int_{B_{2(R+1)}(y)}\frac{1}{2\pi |x-y|}\dd x =2R+2\,.
\]
Using again Lemma~\ref{decaypsi} and the fact that $R\geq 1$ we
can estimate the second integral
\[
\int_{B_{R+1}^c} \frac{1}{2\pi |x-y|} \sqrt{\frac{(R+1)}{|x|}}
\mathrm{e}^{-\kappa (|x|-(R+1))}\dd x \leq
C_5\kappa^{-3/2}\sqrt{R}\mathrm{e}^{\kappa R }\,,
\]
where $C_5=2\int^{\infty}_{0}\sqrt{x}\mathrm{e}^{-x}\dd x$. By
Corollary \ref{c-logdis} there exist constants $\cG_3 $ and $\cG_4
$ such that the estimate
\begin{equation} \label{e-logest}
\int_\Gamma \frac{\alpha}{2\pi |x-y|}\dd \sigma_\Gamma(x) \le
\alpha \l(\cG_3 + \cG_4 (\log L +|\log d_S(y)|)\r)\,
\end{equation}
is valid. Combining the above estimates we obtain the claim.
\end{proof}
\subsection{Lower bound on the gradient}
In Proposition~\ref{p-maximazeros} we have localized the zeros,
minima and maxima of eigenfunctions of $\Hag$. Choose two points
$v_0,v_1 \in \Gamma$ such that
\[
\psi_1(v_0)=\inf_{x \in \RR^2} \psi_1(x) < 0 \quad \text{ and }
\quad \psi_1(v_1)= \sup_{x \in \RR^2} \psi_1(x) > 0 \,,
\]
Taking appropriate scalar multiples
we may assume the normalization $\psi_1(v_1)= \|\psi_1\|_\infty=1$
and  $\|\psi_0\|_\infty =1$.
This means that $f(v_0)<0$ and moreover
\[
f(v_1)=\frac{\|\psi_1\|_\infty}{\psi_0(v_1)} \ge
\frac{\|\psi_1\|_\infty}{\|\psi_0\|_\infty} =1\,.
\]
The following lemma states a lower bound on $\int_{B_R} |\nabla
f(x)| \dd x $.
\begin{prp}\label{p-intgrad}
There exists a positive constant $\beta_0$ such that
\[
\int_{B_R} |\nabla f(y)| \dd y \ge \frac{(\kappa_0\rho)^2 \rho
}{(\alpha \cG_5 +1)^6 (\kappa_0^2 +1)^4\zeta(\kappa
_0)}\mathrm{e}^{-\beta_0 \rho}\,,
\]
where
\[
\zeta(\kappa_0) :=\Big \{ \begin{array}{cc} {1 \,\quad \quad \quad
\mathrm{for }\quad \kappa_0 \geq \frac{1}{2}}
\\ {- \frac{\log \kappa_0}{\log 2} \quad \mathrm{for }\quad \kappa_0 < \frac{1}{2}}
\end{array}
, \quad \quad \rho =\kappa_0 R\,,
\]
$\cG_5=\max\{2\cG_4,\cG_3+\cG_4 \log(\max\{1,L\})\} $ and $\cG_3$
and $\cG_4$ are taken from Corollary~\ref{c-logdis}.
\end{prp}
\begin{proof}
As was already mentioned, in order to estimate $\nabla f$ we will
relay on Proposition ~\ref{p-estgrad} which gives upper bounds on
$\nabla \psi_0$ and $\nabla \psi_1$. Recall that $E_0 =-\kappa
_0^2$, $E_1 =-\kappa_1^2$ are eigenvalues corresponding to
$\psi_0$, respectively $\psi_1$. First let us note that since
$\cS_{\kappa ,\Gamma }$ is an increasing function of $\kappa $ and
$\kappa_0>\kappa_1$ the inequality \eqref{e-estgrad1} implies
\begin{equation} \label{e-estgrad2}
|\nabla \psi_i (y)| \leq \left [\cS+\alpha \cG_4 |\log d_S(y)|
\right]\|\psi_i\|_\infty \,, \quad \mathrm{for }\,\,\, i=0,1\,,
\end{equation}
where we abbreviate
\[
\cS:=\cS_{\kappa_0 , \Gamma }=2\kappa^2(R+1)+C_5
\rho^{1/2}\mathrm{e}^{\rho}+\alpha \l(\cG_3+\cG_4 \log(\max\{1,
L\})\r)\, .
\]
To make use of the inequality \eqref{e-leqgrad} we need some
estimates for $\psi_0^{-1}$ which, in fact, can be directly
derived from Lemma~\ref{l-boundgrst}, i.e. we have
\begin{equation}
\label{e-invground} \sup_{y\in B_R}\psi_0^{-1}(y) \le
C_1^{-1}\kappa_0^{-1}(1+\sqrt{2\rho})\mathrm{e}^{2\rho }
\|\psi_0\|^{-1}\,.
\end{equation}
Combining this with the statement of Lemma~\ref{lem-estnormpsi}
and using our normalization $\|\psi_0\|_\infty =\|\psi_1\|_\infty
=1$ we get
\[
\sup_{y\in B_R} \psi_0^{-1}(y) \leq TD \,,
\]
where
\[
T:= C_1^{-1}\frac{(\cG_5 \alpha +1)^2(\kappa^{2}_{0}+1)}{\kappa_0
\rho}\,, \ D:=(1+\sqrt{2\rho})\mathrm{e}^{(\eta_0+2)\rho} \,,
\]
and $\cG_5 =\max\{2\cG_4,\cG_3+\cG_4(\log\max\{1,L\}) \}$.
Applying the above inequalities to (\ref{e-leqgrad}) and using
again our normalization we have
\begin{eqnarray} \label{e-estf}
|\nabla f (y)|\leq (\cS+\alpha \cG_4 |\log d_S(y)|)TD(TD+1)\,.
\end{eqnarray}
Now choose two parallel line segments $S_0$ and $S_1$, which are
not tangential to $\Gamma$, of length $2b$ and such that $S_i\cap
\Gamma = v_i$ is the midpoint of $S_i$ for $i=0,1$. Thus any $y\in
S_i$ is a $S_i$-admissible point and the expression $d_{S_i}(y)$
is well defined.
%Moreover, we have two angles $\theta_i$, $i=0,1$
%associated with $y\in S_i$ in the sense of
%Definition~\ref{d-admissible}. Due to the above choice we have
%$|\cos \theta_i |\le 1/\sqrt{2}$.

We can suppose without loss of generality, that the line passing
through $v_0$ and $v_1$ is the $y_1$-coordinate axis, $v_0=(0,0)$
and $\cL=\dist(v_0,v_1)$, in other words $v_1=(\cL,0)$.
Furthermore, denote by $\alp$ the angle between the $y_1$-axis and
the segment $S_0$. Let us note that it is always possible to
choose the segments $S_i$in such a way that the smallest angle
formed with the tangential vectors of $\Gamma$ at the points $v_i$
are at least $\pi/6$ and simultanously $\alp$ is also at least
$\pi /6$.

Our first task is to estimate the behavior of $f$ near $v_0$. With
the parameterization assumed above any $y\in S_0$ thus has coordinates
$y=(y_1, y_2 )$ where $y_2=y_1 \tan\alp= d_{S_0}(y) \sin\alp$. For
such $y$ we obtain using the fundamental theorem of calculus and
inequality (\ref{e-estf})
\begin{multline}
\label{e-difference} |f(y) -f(v_0)| =\l|\int_0^{d_{S_0}(y)} \nabla
f(\tau \cos \alp,\tau \sin \alp) \cdot \binom{\cos\alp}{\sin\alp}
\dd \tau \r|
\\
\leq\int_0^{d_{S_0}(y)}  |\nabla f(\tau \cos \alp,\tau \sin
\alp)|\dd \tau \leq \xi (b, \kappa_0, \rho )\,,
\end{multline}
where $\xi (b, \kappa_0, \rho):= b(\cS+\alpha \cG_4 (|\log b
|+1))TD(TD+1)$. Choosing $b$ small enough we can make
$\xi(b,\kappa_0,\rho)$ arbitrarily small. More precisely, by
Lemma~\ref{l-xicov} and Corollary~\ref{c-xicov}  we know that
there exists a positive constant $\beta_0$ such that
\begin{equation}
\label{e-1/4} \xi (b, \kappa_0, \rho ) \leq \frac{1}{4}\quad
\mathrm{for }\quad b=\frac{(\kappa_0\rho)^2 \rho }{(\alpha \cG_5
+1)^6 (\kappa_0^2 +1)^4\zeta(\kappa_0)}\mathrm{e}^{-\beta_0
\rho}\,, \end{equation} where the function $\zeta$ is defined in
the statement of the proposition.

Finally, using (\ref{e-difference}) we get
\[
f(y)\leq \frac{1}{4}\quad \text{ on } \quad S_0 \,.
\]
Similarly, for $S_1$, respectively $b$ small enough we obtain
$f(y) \ge 3/4$ for all $y\in S_1$. Using these inequalities we
estimate the integral of the gradient of $f$ on the strip $\cT:=
\{(y_1,y_2) \in \RR^2\mid y_1 = \tau \cos\alp+l,y_2 =\tau\sin
\alp, \tau \in [-b,b], l\in [0,\cL]\}$
\begin{multline*}
\int_\cT |\nabla f(y)| \dd y \ge \int_\cT |\partial_{y_1} f(y)|
\dd y
\\
\geq |\sin \alp \int_{-b}^b \l(f(\tau \cos\alp+\cL,\tau\sin
\alp)-f(\tau \cos\alp,\tau\sin \alp)\r)\dd \tau| \geq  |\sin \alp|
b\,,
\end{multline*}
where we again employ the fundamental theorem of calculus. Using
the fact that $|\sin \alp|\geq 1/2$ we get
\begin{equation}\label{e-estnabla}
\int_{B_R} |\nabla f(y)| \dd y \ge \frac{(\kappa_0\rho)^2
\rho}{2(\alpha \cG_5 +1)^6 (\kappa_0^2 +1)^4\zeta(\kappa
_0)}\mathrm{e}^{-\beta_0 \rho}\,.
\end{equation}

\end{proof}

\begin{proof}[Proof of Theorem~\ref{t-spgaps}]
Inserting the inequalities given in Lemmata~\ref{l-boundgrst},
\ref{lem-norms}, \ref{lem-estnormpsi} and \ref{p-intgrad} into the
estimate (\ref{1estim}) yields
\[
E_0 -E_1 \geq \kappa_1^2 \,\mu_{\Gamma,\alpha}(\rho ,\kappa_0)\,
\mathrm{e}^{-(8+2\eta_0+ 2\beta_0)\rho}\,,
\]
where
\begin{equation}\label{defmu}
\mu_{\Gamma,\alpha}(\rho ,\kappa_0)
:=C_7\frac{(\kappa_0\rho)^8}{(L\alpha)^2 (1+\sqrt{2\rho})^4
(\kappa_0^2+1)^{10}(\cG_5 \alpha +1 )^{16}\xi (\kappa_0)^2}\,,
\end{equation}
where $C_7$ is an absolute constant,
$\cG_5:=\max\{2\cG_4,\cG_3+\cG_4 \log(\max\{1,L\})\} $ and $\cG_3$
and $\cG_4$ are taken from Corollary~\ref{c-logdis}. This proves
the claim.
\end{proof}

\section{Closing remarks and open questions}
\label{s-Outlook}
\subsubsection*{Dependence on the second eigenvalue.}
Apart from the parameter $\kappa _0$ corresponding to the ground
state energy, $\kappa_1$ is also involved in the lower bound for
the first spectral gap
\begin{equation}\label{mainest}
E_1 -E_0 \geq \kappa_1^2 \mu_{\Gamma ,\alpha }(\rho ,
\kappa_0)\mathrm{e}^{-C_0\rho}\,, \text{ with } \rho:=
\kappa_0R\,.
\end{equation}
In fact, the appearance of $\kappa_1$ here is natural since we
assumed explicitely the existence of the second, isolated
eigenvalue. For effective estimates of the spectral gap we would
need lower bounds on $\kappa_1^2$. The following observation is
helpful in many situations. Suppose that for a Hamiltonian $\Hag$
the value $C_{\alpha ,\gamma}$ is a lower bound for $\kappa_1^2$.
Then $C_{\alpha ,\gamma}$ is also the corresponding lower bound
for all $H_{\tilde{\alpha} ,\sigma_{\tilde{\Gamma}}}$, where
$\tilde{\alpha}\geq \alpha$ and  $\tilde{\Gamma}\supset\Gamma$.
The above statement is a direct consequence of the form sum
representation of the Hamiltonian and the min-max theorem. In
general the question whether a second eigenvalue exists is quite
involving and will be discussed elsewhere.

\bigskip

\subsubsection*{Strong coupling constant case.} There is one case
where the function counting the number of eigenvalues is known.
This is the situation where the singular interaction is very
strong, more precisely $\alpha \to \infty $, cf.
Remark~\ref{r-spectrum}. In particular, if $\Gamma $ consists of one
closed curve, the asymptotic behaviour of the eigenvalues
implies the following estimate on the first spectral gap
\[
E_{1} -E_0 =\mu_{1} -\mu_0 +\cO\left( \frac{\log \alpha
}{\alpha }\right) \quad\mathrm{ for }\quad \alpha \to \infty\,,
\]
where $\mu_0,\mu_1$ are the two lowest eigenvalues of an
appropriate comparison operator. This operator is defined as the
negative Laplacian with periodic boundary conditions on $[0,L]$
plus a regular potential. Therefore, the estimate on the first
spectral gap for $\Hag$ can be expressed by
the %corresponding
gap for a Schr\"{o}dinger operator with an ordinary potential.
Denote by $\phi_0$ the ground state of the comparison operator
and
\[
a := \left( \frac{\max _{x\in [0,L]}\phi_0(x)}{\min _{x\in [0,L]}\phi_0(x)}\right)^2 \, .
\]
Theorem 1.4 in \cite{KirschS-87} implies
\[
a^{-1} \left( \frac{2\pi}{L}\right)^2
\le \mu_1 - \mu_0
\le a\left( \frac{2\pi}{L}\right)^2.
\]
Note that the quotient $a$ is independent of scaling by $L$. Since
we are considering a single closed curve, $L/2 < R$ and thus
$\mu_1 - \mu_0 \ge a^{-1} \left( \frac{\pi}{R}\right)^2 $. Hence
in the considered situation the lowest spectral gap decreases only
polynomially in $1/R$, rather than exponentially as estimated in
Theorem \ref{t-spgaps}.

\smallskip

\subsubsection*{Singular perturbation on an infinite curve.}
It is very natural to pose the question whether the results
obtained in the present paper can be extended to Hamiltonians
with a singular potential supported on an infinite curve.
If the curve is asymptotically straight in an appropriate sense
then the essential spectrum is the same as in the case of a
straight line. If the curve is non-straight, the existence of
at least one isolated eigenvalue was shown in \cite{ExnerI-01}
and the function counting the number of eigenvalues
for the strong coupling constant case was derived in \cite{ExnerY-02b}.

The estimate obtained in the main Theorem~\ref{t-spspegap}
should hold for infinite curves as well.
However, to obtain this result, one has to analyse the behavior of
certain eigenfunctions and prove their exponential decay,
a problem which is not encountered in the case of a finite curve.
We postpone this question  to a subsequent publication.

\bigskip

Related estimates about eigenvalue splittings for certain infinite
quantum waveguides have been derived in \cite{BorisovE-04}.

\appendix
\section{Proof of inequality \eqref{e-1/4}}
\label{s-appendix}

In this appendix we complete a technical estimate which is needed
in the proof of Proposition \ref{p-intgrad}. More precisely, we
prove here the inequality \eqref{e-1/4} on the function
$\xi(b,\kappa_0,\rho)$.

For the reader's convenience let us recall some notation
introduced in the proof of Proposition~\ref{p-intgrad}. Define
\[
\xi (b, \kappa_0, \rho):= b(\cS+\alpha \cG_4 (|\log b
|+1))TD(TD+1)\,,
\]
where
\[
T:= C_1^{-1}\frac{(\cG_5 \alpha +1)^2(\kappa^{2}_{0}+1)}{\kappa_0
\rho}\,, \,\, D:=(1+\sqrt{2\rho})\mathrm{e}^{(\eta_0+2)\rho} \,
\]

$\rho =\kappa_0 R$ and
$\cS:=\kappa^2_0(2R+2)+C_5\rho^{1/2}\mathrm{e}^{\rho}+\alpha
(\cG_3+\cG_4\log(\max\{1,L\}))$. Let us introduce a one parameter
family of functions defined by
\[
b_\beta (\rho , \kappa_0 )=\frac{(\kappa_0\rho)^2 \rho }{(\alpha
\cG_5 +1)^6 (\kappa_0^2 +1)^4\zeta(\kappa_0) }\mathrm{e}^{-\beta
\rho}\,,\quad \zeta (\kappa_0 ):=\Big \{
\begin{array}{ll}
1 & \text{ for } \kappa_0 \geq \frac{1}{2} \\
-\frac{\log \kappa_0}{\log 2} & \text{ for } \kappa_0 <\frac{1}{2}\,.
\end{array}
\]

\begin{lem}\label{l-xicov}
There exists an absolute constant $C_{9}$ such that for $\beta >
2\eta_0 +5$ we have
\[
\xi (b_\beta ,\kappa_0 ,\rho )\leq \frac{C_{9}}{\beta -2\eta_0
-5}\, ,
\]
uniformly in $\rho$ and $\kappa_0$.
\end{lem}
\begin{proof}
In the following proof we will use the fact that the terms in the
enumerator of $T$ as well as $\zeta (\kappa_0)$ are larger or
equal $1$. Using the formula for $T$ we obtain
\begin{eqnarray}\nonumber \xi (b, \kappa_0, \rho
)\leq b T^2(\cS+\alpha \cG_4 (|\log b |+1))D(D+\kappa_0\rho )\\
\label{e-estxi1} \leq b T^2(\cS+2\alpha \cG_4 |\log b
|)D(D+\kappa_0\rho ) \,,
\end{eqnarray}
where in the last inequality we assume that $b<e^{-1}$.
Furthermore applying the explicit form for $\cS$ we get by a
straightforward calculation that the right hand side of
(\ref{e-estxi1}) is bounded from above by
\begin{multline}
b T^2(\kappa_0^2 +1)^2 (\alpha \cG_5 +1)^2\zeta (\kappa_0 )
\\ \label{e-estxi2}
\Big (3+2\rho +C_5\rho^{1/2}\mathrm{e}^\rho +\frac{|\log b|}{
(\kappa_0^2 +1) (\alpha \cG_5 +1)\zeta (\kappa_0 )}\Big )
D(D+\rho)\,.
\end{multline}

Employing the definition of $b_\beta $ and inserting this in the
expression (\ref{e-estxi2}) we get that (\ref{e-estxi2}) is
smaller or equal to
\begin{equation}\label{e-estxi3}
C_1^{-2} \rho \mathrm{e}^{-\beta \rho }\Big (3+2\rho
+C_5\rho^{1/2}\mathrm{e}^\rho + \frac{|\log b_\beta |}{(\kappa_0^2
+1) (\alpha \cG_5 +1)\zeta (\kappa_0 )} \Big) D(D+\rho)\,.
\end{equation}
Let us estimate now the logarithmic term in (\ref{e-estxi3}).
Using again the formula for $b_\beta $ and properties of the
logarithmic function we get
\begin{multline*}
\frac{|\log b_\beta |} { (\kappa_0^2 +1) (\alpha \cG_5 +1)\zeta
(\kappa_0 )}
\\
\leq \frac{|\log \zeta (\kappa_0)|}{\zeta (\kappa_0)}+2\frac{|\log
\kappa_0|}{\zeta (\kappa_0) (\kappa_0^2 +1) }+3|\log \rho | +\beta
\rho + \frac{\log\Big ( (\kappa_0^2 +1)^4 (\alpha \cG_5
+1)^6\Big)}{(\kappa_0^2 +1) (\alpha \cG_5 +1)}
\\
\leq C_8 +3|\log \rho |+\beta \rho \,.
\end{multline*}
In the last inequality we estimated the first, second and last
term by a constant $C_8$. Inserting this to (\ref{e-estxi3}) we
obtain that (\ref{e-estxi3}) is bounded from above by
\begin{equation}\label{e-estxi4}
C_1^2\rho \mathrm{e}^{-\beta \rho }(C_8 + 2\rho
+C_5\rho^{1/2}\mathrm{e}^\rho +3|\log \rho |+\beta \rho
)D(D+\rho)\,.
\end{equation}
Employing now the explicit form for $D$ we estimate
\begin{eqnarray}
\nonumber \text{expression } (\ref{e-estxi3})\,\leq \Xi_\beta
(\rho )\,,
\end{eqnarray}
where
\[
\Xi_\beta (\rho ):= C_1^2\rho \nonumber \mathrm{e}^{(-\beta
+2\eta_0 +5) \rho }(C_8+ 2\rho +C_5\rho^{1/2} +3|\log \rho |+\beta
\rho )(1+\sqrt{2\rho})(1+\sqrt{2\rho}+\rho)\,.
\]
Now we estimate the maximum of the functions $\rho \mapsto
\Xi_\beta(\rho) $ and conclude that there exists a positive
constant $C_{9}$ such that
\[
\Xi_\beta (\rho )\leq \frac{C_{9}}{\beta -2\eta_0 -5}\,,
\]
for $\beta > 2\eta_0 +5$. This proves the desired claim.
\end{proof}

\begin{cor} \label{c-xicov}There exists constant $\beta_0$ such that for any
$\beta \geq \beta_0$ we have
\[
\xi (b_\beta ,\kappa_0 ,\rho )\le \frac{1}{4}\,,
\]
 for all values of $\kappa_0 ,\rho$.
\end{cor}

Of course, $\beta_0$ should be chosen in such a way that
$b_{\beta_0}\leq b_1$, where $b_1$ is defined in
Lemma~\ref{l-logdis}. This is always possible because
\[
b_\beta (\rho , \kappa_0)\leq \rho^3 \mathrm{e}^{-\beta\rho}\leq
9\mathrm{e}^{-3} \beta^{-3}\,.
\]
%In particular there exists a value $\beta_2$ such
%that for $\beta \ge \beta_2$ we have $\xi (b_\beta ,\kappa_0 ,\rho
%)\le \frac{1}{4}$ for all values of $\kappa_0 ,\rho$. Setting
%$b_2:= b_{\beta_2}(\rho,\kappa_0)$ we have
%\[
%\xi (b_2 ,\kappa_0 ,\rho )\le \frac{1}{4}\,.
%\]

% \bibliographystyle{alpha}
% \bibliography{ILit-05}

\begin{thebibliography}{BEK{\v{S}}94}

\bibitem[Agm85]{Agmon-85}
Sh. Agmon.
\newblock Bounds on exponential decay of eigenfunctions of {S}chr\"odinger
  operators.
\newblock In {\em Schr\"odinger operators (Como, 1984)}, volume 1159 of {\em
  Lecture Notes in Math.}, pages 1--38. Springer, Berlin, 1985.

\bibitem[AS72]{AbramowitzS-72}
M.~Abramowitz and I.~A. Stegun.
\newblock {\em Handbook of Mathematical Functions}.
\newblock Dover, New York, 1972).

\bibitem[BE04]{BorisovE-04}
D.~Borisov and P.~Exner.
\newblock Exponential splitting of bound states in a waveguide with a pair of
  distant windows.
\newblock {\em J. Phys. A}, 37(10):3411--3428, 2004.

\bibitem[BEK{\v{S}}94]{BrascheNEKS-94}
J.~F. Brasche, P.~Exner, Yu.~A. Kuperin, and P.~{\v{S}}eba.
\newblock Schr\"odinger operators with singular interactions.
\newblock {\em J. Math. Anal. Appl.}, 184(1):112--139, 1994.

\bibitem[DS84]{DaviesS-84}
E.~B. Davies and B.~Simon.
\newblock Ultracontractivity and the heat kernel for {S}chr\"odinger operators
  and {D}irichlet {L}aplacians.
\newblock {\em J. Funct. Anal.}, 59(2):335--395, 1984.

\bibitem[EI01]{ExnerI-01}
P.~Exner and T.~Ichinose.
\newblock Geometrically induced spectrum in curved leaky wires.
\newblock {\em J. Phys. A}, 34(7):1439--1450, 2001.

\bibitem[EK02]{ExnerK-02}
P.~Exner and S.~Kondej.
\newblock Curvature-induced bound states for a {$\delta$} interaction supported
  by a curve in {$\mathbb R\sp 3$}.
\newblock {\em Ann. Henri Poincar\'e}, 3(5):967--981, 2002.

\bibitem[EK03]{ExnerK-03}
P.~Exner and S.~Kondej.
\newblock Bound states due to a strong {$\delta$} interaction supported by a
  curved surface.
\newblock {\em J. Phys. A}, 36(2):443--457, 2003.

\bibitem[Exn03]{Exner-03}
P.~Exner.
\newblock Spectral properties of {S}chr\"odinger operators with a strongly
  attractive {$\delta$} interaction supported by a surface.
\newblock In {\em Waves in periodic and random media (South Hadley, MA, 2002)},
  volume 339 of {\em Contemp. Math.}, pages 25--36. Amer. Math. Soc.,
  Providence, RI, 2003.

\bibitem[Exn05]{Exner-05}
P.~Exner.
\newblock An isoperimetric problem for leaky loops and related mean-chord
  inequalities.
\newblock {\em J. Math. Phys.}, 46(6):062105, 2005.
\newblock http://arxiv.org/abs/math-ph/0501066.

\bibitem[EY02]{ExnerY-02b}
P.~Exner and K.~Yoshitomi.
\newblock Asymptotics of eigenvalues of the {S}chr\"odinger operator with a
  strong {$\delta$}-interaction on a loop.
\newblock {\em J. Geom. Phys.}, 41(4):344--358, 2002.

\bibitem[EY03]{ExnerY-03}
P.~Exner and K.~Yoshitomi.
\newblock Eigenvalue asymptotics for the {S}chr\"odinger operator with a
  {$\delta$}-interaction on a punctured surface.
\newblock {\em Lett. Math. Phys.}, 65(1):19--26, 2003.

\bibitem[EY04]{ExnerY-04}
P.~Exner and K.~Yoshitomi.
\newblock Erratum: ``{E}igenvalue asymptotics for the {S}chr\"odinger operator
  with a {$\delta$}-interaction on a punctured surface'' [{L}ett.\ {M}ath.\
  {P}hys.\ {\bf 65} (2003), no.\ 1, 19--26].
\newblock {\em Lett. Math. Phys.}, 67(1):81--82, 2004.

\bibitem[GT83]{GilbargT-83}
D.~Gilbarg and N.~S. Trudinger.
\newblock {\em Elliptic partial differential equations of second order}.
\newblock Springer, Berlin, 1983.

\bibitem[Har80]{Harrell-80}
E.~M. Harrell.
\newblock Double wells.
\newblock {\em Comm. Math. Phys.}, 75(3):239--261, 1980.

\bibitem[KS85]{KirschS-85}
W.~Kirsch and B.~Simon.
\newblock Universal lower bounds of eigenvalue splittings for one dimensional
  {Schr\"odinger} operators.
\newblock {\em Commun. Math. Phys.}, 97:453--460, 1985.

\bibitem[KS87]{KirschS-87}
W.~Kirsch and B.~Simon.
\newblock Comparison theorems for the gap of {Schr\"odinger} operators.
\newblock {\em J. Funct. Anal.}, 75:396--410, 1987.

\bibitem[Pos01]{Posilicano-01}
A.~Posilicano.
\newblock A {K}re\u\i n-like formula for singular perturbations of self-adjoint
  operators and applications.
\newblock {\em J. Funct. Anal.}, 183(1):109--147, 2001.

\bibitem[Pos04]{Posilicano-04}
A.~Posilicano.
\newblock Boundary triples and {W}eyl functions for singular perturbations of
  self-adjoint operators.
\newblock {\em Methods Funct. Anal. Topology}, 10(2):57--63, 2004.

\end{thebibliography}

\def\cprime{$'$} \def\cprime{$'$}

\end{document}